\documentclass[pdflatex,sn-mathphys-num]{sn-jnl}% Math and Physical Sciences Numbered Reference Style
\usepackage{graphicx}%
\usepackage{multirow}%
\usepackage{amsmath,amssymb,amsfonts}%
\usepackage{amsthm}%
\usepackage{mathrsfs}%
\usepackage[title]{appendix}%
\usepackage{xcolor}%
\usepackage{textcomp}%
\usepackage{manyfoot}%
\usepackage{booktabs}%
\usepackage{algorithm}%
\usepackage{algorithmicx}%
\usepackage{algpseudocode}%
\usepackage{listings}%
\usepackage{subfigure}

\usepackage{tikz}
\usepackage{tabularx}
\usetikzlibrary{shapes.geometric, arrows}
\tikzstyle{process} = [rectangle, minimum width=3cm, minimum height=1cm, text centered, draw=black] %fill=orange!30]

\theoremstyle{thmstyleone}%
\theoremstyle{thmstyletwo}%

\theoremstyle{thmstylethree}%

\begin{document}

\title[Potential Effects of Loading Terminal Locations on Surface Trajectories of Oil Transport]{Potential Effects of Loading Terminal Locations on Surface Trajectories of Oil Spill Transport}


\author[1]{\fnm{Shoshana}  \sur{Reich}} %\email{sreich@utexas.edu}
%I am not sure if you submit with Shoshi or Shoshana 

\author[2]{\fnm{Edward} \sur{Buskey}} %\email{ed.buskey@utexas.edu}

\author[1]{\fnm{Clint} \sur{Dawson}} %\email{clint.dawson@oden.utexas.edu}

\author*[3,4]{\fnm{Eirik} \sur{Valseth}}\email{eirik.valseth@nmbu.no}

\affil*[3]{\orgdiv{Department of Data Science, Faculty of Science and Technology}, \orgname{Norwegian University of Life Sciences}, \orgaddress{\street{Drøbaksveien 31}, \city{Ås}, \postcode{1433}, \country{Norway}}}

\affil[4]{\orgdiv{Department of Numerical Analysis and Scientific Computing}, \orgname{Simula Research Laboratory}, \orgaddress{\street{Kristian Augusts Gate 23}, \city{Oslo}, \postcode{0164}, \country{Norway}}}

\affil[1]{\orgdiv{The Oden Institute for Computational Sciences and Engineering}, \orgname{The University of Texas at Austin}, \orgaddress{\street{201 E. 24th St. Stop C0200}, \city{Austin}, \postcode{78712}, \state{Texas}, \country{USA}}}

\affil[2]{\orgdiv{Marine Sciences Institute}, \orgname{The University of Texas at Austin}, \orgaddress{\street{750 Channel View Drive}, \city{Port Aransas}, \postcode{78373}, \state{Texas}, \country{USA}}}


\abstract{We present an investigation comparing the potential impacts of offshore and onshore crude oil loading sites on surface trajectories of spilled oil particles in the regions near the Port of Corpus Christi, Texas. 
Oil transport is established in a two step procedure. First, the circulation and flow characteristics of seawater throughout the coastal ocean are established for various flow conditions, including current and proposed channel depth, seasonality changes, and extreme weather events. Then, spilled oil is modeled as distinct particles released at either the proposed onshore or offshore loading locations. The particle trajectories are tracked and used to assess the spread into diverse coastal ecosystems with extensive plant, sea, and land life. The models indicate that the extent of spread of these simulated oil spills to ecologically significant regions is greater when initiated at the onshore loading site than at the offshore site. }

%%================================%%
%% Sample for structured abstract %%
%%================================%%

\keywords{Hydrodynamics, ADCIRC, Lagrangian Particle Tracking}



\maketitle
\section{Introduction}

\begin{figure}[h!]
\centering
 \includegraphics[width=0.75\textwidth]{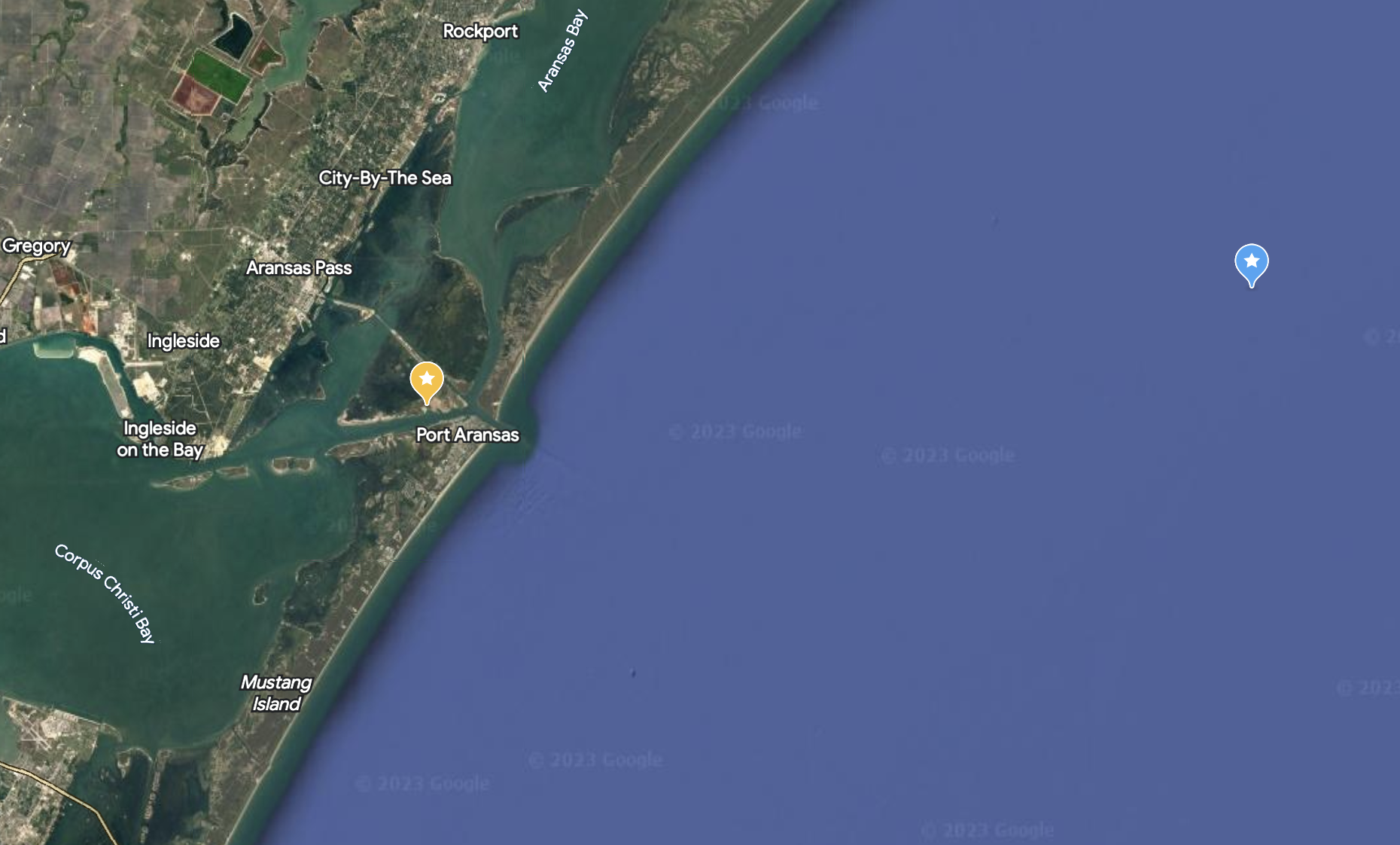}
  \caption{\label{fig:terminal_locations} Locations of the proposed loading terminals. The Harbor Island onshore deepwater dock is shown in yellow, and the offshore Bluewater Texas Terminal is shown in blue. Satellite photo obtained from Google Earth. }
\end{figure}

The Port of Corpus Christi accounts for approximately 60\% of all U.S. crude oil exports by volume\protect\footnote{https://portofcc.com/port-of-corpus-christi-finishes-fiscal-year-2022-with-record-tonnage/}.
There are several proposed industrial developments in the region associated with oil exports, including new offshore and onshore tanker loading locations. The proposed onshore deepwater dock % U.S.A.C.E.% 
on Harbor Island would be capable of docking and fully loading two Very Large Crude Carriers (VLCCs), which are 330m long vessels that can carry up to two million barrels of oil. The proposed offshore Bluewater Texas Terminal (BWTX) would be located 21 miles east of the Port of Corpus Christi and capable of loading 1.56 million barrels per day. Figure~\ref{fig:terminal_locations} shows the locations of the two proposed sites.

\begin{figure}[h!]
\centering
 \includegraphics[width=0.75\textwidth]{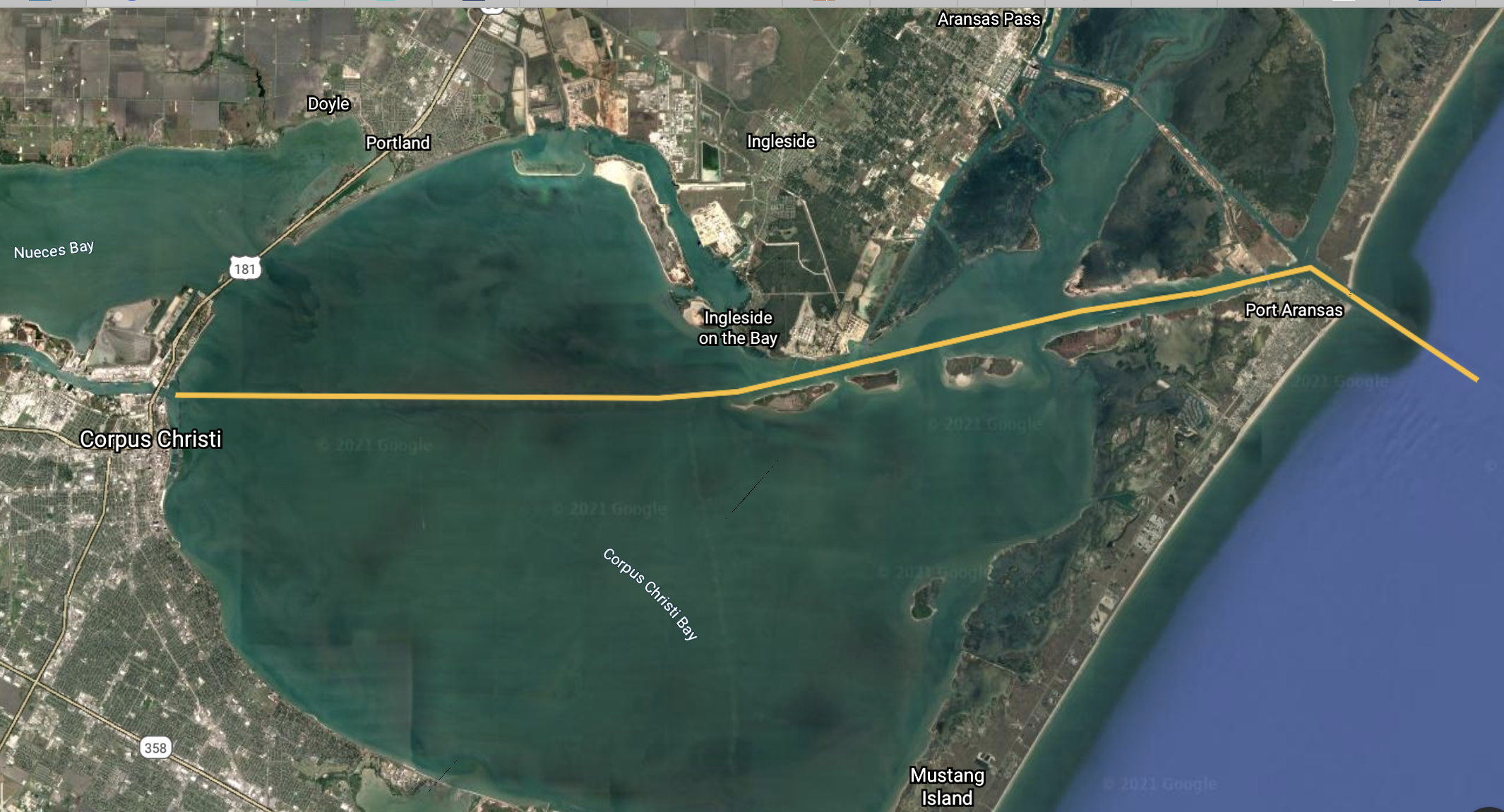}
  \caption{\label{fig:channel} Location of the Corpus Christi ship channel. Satellite photo obtained from Google Earth. }
\end{figure}

To accommodate the expansion of crude oil exports and larger ships in the Port of Corpus Christi, it has been proposed to deepen the section of  the Corpus Christi ship channel between the Gulf of Mexico and Harbor Island, Port 
from an average depth of $14.33m$ ($47ft$) to an average depth of $21.33m$ ($70ft$)~\cite{hamilton2018alternatives}. The channel connects the Gulf to Corpus Christi Bay and Aransas Pass~\cite{hamilton2018alternatives} (see Figure~\ref{fig:channel}), and it allows ships to enter and exit the system of bays in order to reach the port through the pass. Along the ship channel, through Corpus Christi Bay and Nueces Bay there are numerous areas that are home to a multitude of diverse animal and plant species, including Red Drum fish and their larvae~\cite{valseth2021study}, as well as birds such as Sandhill and Whooping Crane~\cite{mclean2019survival} and other land-based species such as deers in the Aransas Wildlife Refuge~\cite{halloran1943management} . %Due to the growth and expansion of the Port of Corpus Christi, it has been proposed to deepen the ship channel through the pass to  to accommodate larger ships at the port. 
The proposed channel depth and subsequent change in bottom topography will lead to changes in the flow characteristics of the water through Aransas Pass. In previous studies, we considered the effects of a deepened channel on hurricane storm surge~\cite{valseth2022study} and on the transport of Red Drum fish larvae~\cite{valseth2021study}. These studies indicate that the flows and hydrodynamics through the Aransas Pass changed, however, the impacts of these changes were minor.

In this work, we compare the effects of the deeper channel, seasonal changes, and extreme weather events on surface oil trajectories at the proposed onshore and offshore loading locations. We developed a model governing the water circulation using both current and proposed channel depths at the onshore and offshore loading locations at several times of year. From these circulation models, we ascertain the velocity components throughout the modeled region. Oil spills are modeled as injections of distinct particles with passive transport resulting from the water circulation.

In the following, the modeling methods are described in detail in Sections~\ref{sec:modeling_hyrdo} and~\ref{sec:modeling_particles}. 
In Section~\ref{sec:results}, we present the results from our model and compare findings based on current and proposed channel depths for the proposed onshore and offshore crude oil loading sites. Finally, in Section~\ref{sec:conclusions} we conclude this report with remarks on the findings. The term "bathymetry" is used throughout this document and it refers to the depth of water relative to the North American Vertical Datum of 1988 (NAVD88). Here, we use the convention that the bathymetry is positive, i.e., dry land, above NAVD88.

%- two proposed locations: BWTX (offshore) and VLCCs (onshore)\\
%- Port Aransas, proposed industrial developments associated with the export of oil\\
%- existing and future depths: POCC proposing to deepen the Corpus Christi Ship Channel, to accomodate larger vessels to carry more oil\\
%- Harbor Island VLCC terminal: two onshore deepwater docks proposed in Harbor Island, one could hold two Very Large Crude Carriers (VLCCs) (1100 ft, 2 million barrels) and another to hold two Suezmax-sized vessels (900 ft, 1 million barrels); able to accomodate the full loading of two VLCCs\\
%- two offshore loading terminals: but we looked at the Phillips 66 Bluewater Texas Terminal, fill the carrier without going onshore, 21 miles east of POCC; 1.56 million barrels per hour

\section{Modeling Methodology}
\label{sec:modeling}

To simulate the transport of oil particles, we first determine the flow characteristics of the coastal circulation (here referred to as hydrodynamics) by employing the advanced circulation (ADCIRC) model~\cite{luettich1992adcirc} - a well documented and extensively tested model. ADCIRC solves the governing mathematical equations for coastal circulation, the Shallow Water Equations using the finite element method~\cite{oden2006finite}. This model and some of its capabilities are discussed further in Section~\ref{sec:modeling_hyrdo}.
Subsequently, the hydrodynamic model results are then used as input in a Lagrangian particle transport model to track the trajectories of oil, which is modeled as distinct particles. This model is described further in Section~\ref{sec:modeling_particles}.

\subsection{Hydrodynamics}
\label{sec:modeling_hyrdo}

\begin{figure}[h!]
    \centering
    %\scalebox{.625}{\input{domain_t.tex} }%
    \begin{tikzpicture}

    % Define colors for water and land
    \definecolor{waterblue}{RGB}{150,212,250}
    \definecolor{sand}{RGB}{194,178,178}
    
    % Draw the horizontal datum
    \draw[black, very thick] (-4,0) -- (4,0);
    %\node[right] at (4.5,0) {Horizontal Datum};
    
    % Draw the seabed (variable bathymetry)
    \draw[brown, thick] plot[smooth, tension=.7] coordinates {(-4,-2.5) (-2,-1.8) (0,-2.3) (2,-2.0) (4,-2.4)};
    \fill[sand] plot[smooth, tension=.7] coordinates {(-4,-3) (-4,-2.5) (-2,-1.8) (0,-2.3) (2,-2.0) (4,-2.4) (4,-3)} -- cycle;

    % Draw the water surface (variable)
    \draw[blue, thick] plot[smooth, tension=.7] coordinates {(-4,1.0) (-2,0.8) (0,1.2) (2,1.1) (4,1.3)};
    \fill[waterblue] plot[smooth, tension=.7] coordinates {(-4,1.0) (-2,0.8) (0,1.2) (2,1.1) (4,1.3)} -- plot[smooth, tension=.7] coordinates { (4,-2.4) (2,-2.0) (0,-2.3) (-2,-1.8) (-4,-2.5) } -- cycle;
    
    % Depth line and label at a specific point
    %\draw[dashed] (1, 1.1) -- (1, -2.2);
    \node[right] at (1, -0.6) {$\quad H = \zeta + h_b$};
    
    % Add labels
    \node[right] at (4.5, 0.450) {$\quad \zeta $};
    \node[right] at (4.5, -1.2) {$\quad h_b$};

    % Depth arrow
    \draw[<->] (1.3, 1.1) -- (1.3, -2.1);

    % Arrows and labels for water surface elevation and bathymetry
    \draw[<->] (4.5, 1.3) -- (4.5, 0);
    \node[above] at (4.5, 1.3) { };

    \draw[<->] (4.5, 0) -- (4.5, -2.4);
    \node[below] at (4.5, -2.4) { };
    
    \draw[black , thick] (-4,0) -- (4,0);

\end{tikzpicture}
    \caption{Definition of shallow water elevations. The horizontal line denotes the geoid, where $\zeta = h_b = 0$.}
    \label{fig:elevation_def}
\end{figure}
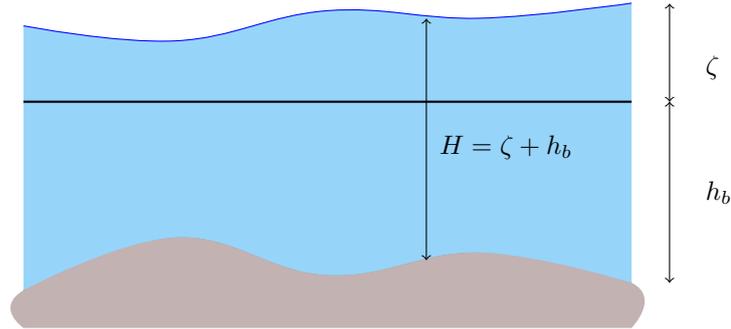

\begin{figure}[h!]
\centering
 \includegraphics[width=0.75\textwidth]{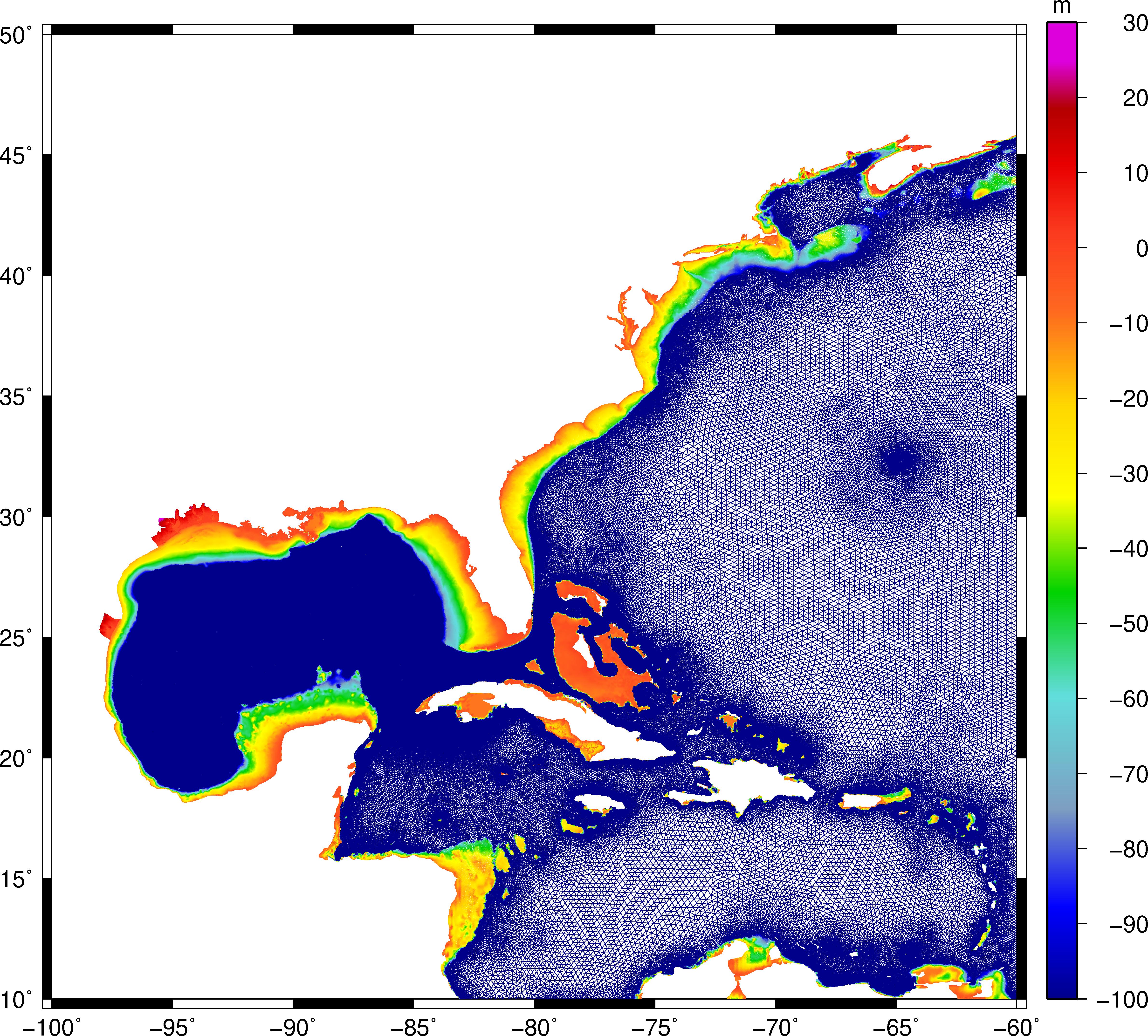}
  \caption{\label{fig:adcirc_mesh} Model domain covered by the ADCIRC finite element mesh. Note that the colorbar denotes the bathymetry measured in meters cut off at $100m$ for the sake of presentation.}
\end{figure}

We model the flow of seawater using the two-dimensional shallow water equations (SWE) for the conservation of mass and momentum~\cite{vreugdenhil2013numerical}. These nonlinear and transient partial differential equations are, in conservative form:
\begin{equation} \label{eq:SWE}
\begin{array}{ll}
\quad \text{Find }  (\zeta, \mathbf{u})   \text{ such that:}  \qquad \qquad    \\ \\
 \frac{\displaystyle \partial  \zeta}{\partial t} + \text{div} (H{\mathbf{u}})  = 0, \text{ in } \Omega, & \\ \\
\frac{ \displaystyle \partial (Hu_x)}{\partial t} + \text{div} \left( Hu_x^2 + \frac{\displaystyle g}{2}(H^2-h_b^2), Hu_xu_y \right) - g\zeta \frac{\displaystyle \partial h_b}{\partial x}  = F_x, \text{ in } \Omega, & \\ \\
\frac{\displaystyle \partial (Hu_y)}{\partial t} + \text{div} \left( Hu_xu_y, Hu_y^2 + \frac{\displaystyle g}{2}(H^2-h_b^2) \right) - g\zeta \frac{\displaystyle \partial h_b}{\partial y}  = F_y, \text{ in } \Omega, 
 \end{array}
\end{equation}
where $\zeta$ is the surface elevation above the North American Vertical Datum of 1988 (NAVD88),   $h_b$ the bathymetry, $H =\zeta + h_b$ is the total water column (see Figure \ref{fig:elevation_def}),  $\mathbf{u} = \{ u_x,u_y\}^{\text{T}}$ is the depth averaged velocity field, and the source terms $F_x,F_y$ represent potential relevant sources which induce flow, including: Coriolis force, tides, wind stresses, and wave radiation stresses, as well as potential viscous effects. Finally, $\Omega$ is the computational domain, see Figure \ref{fig:adcirc_mesh}. 
The solution of these equations over physical domains of interest in this case requires the use of numerical approximations.
Using the ADCIRC model, the SWE are approximated in space using a first order  Bubnov-Galerkin finite element method~\cite{oden2006finite} and a semi implicit time finite difference method in time. To increase the numerical stability in the solution of the SWE~\eqref{eq:SWE}, in ADCIRC, the continuity equation is modified into a form called the generalized wave continuity equation (GWCE)~\cite{lynch1979wave}. Due to the complex and irregular structure of the coastline, the finite element method natural capability to utilize unstructured meshes makes ADCIRC very well suited for the current application. 

%ADCIRC is well suited to this application as it uses unstructured meshes that are capable of accurately description of intricate coastal domains.   

%To develop an ADCIRC model here, we need the following components: $i)$ a discrete description of the domain (i.e., a finite element mesh), including
%bathymetry and bottom friction information, $ii)$  information about tides, and $iii)$ meteorological data. The latter two are the driving mechanisms 
%of the flow. 

\begin{figure}[h]
\subfigure[ \label{fig:bathy_comp1} Existing. ]{\centering
\includegraphics[width=0.5\textwidth]{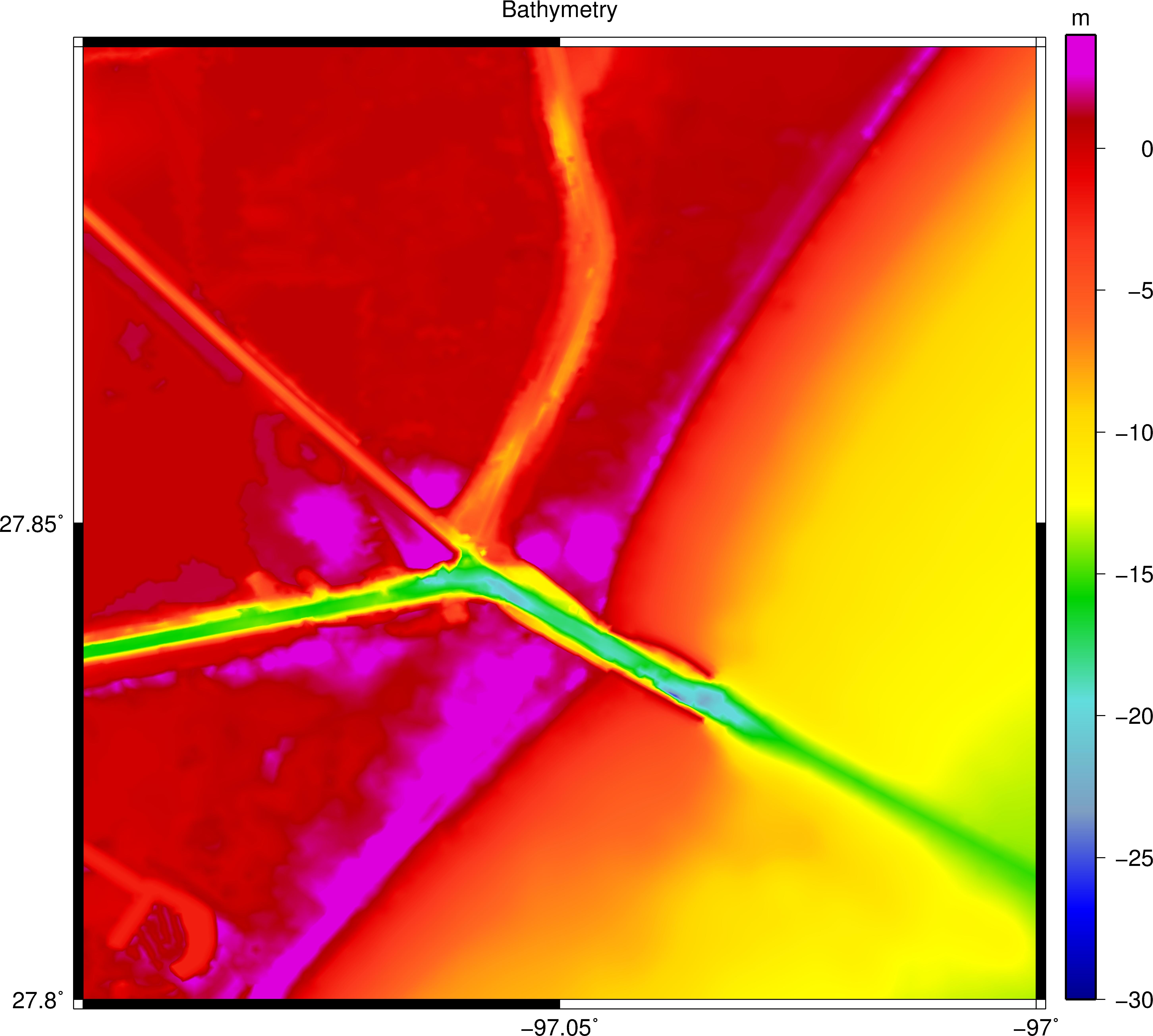}}
\subfigure[ \label{fig:bathy_comp2} Proposed.  ]{\centering
\includegraphics[width=0.5\textwidth]{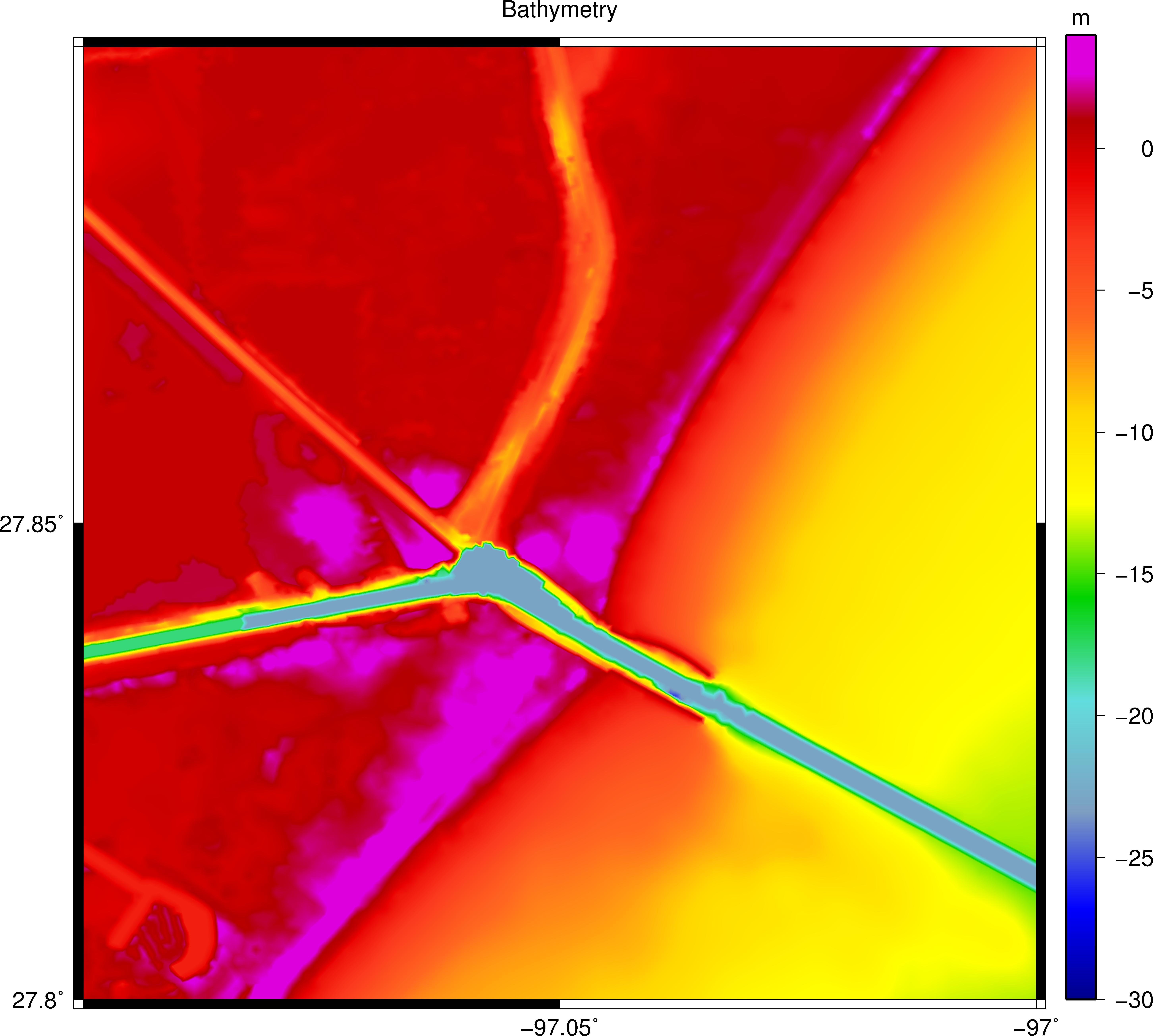}}
\caption{\label{fig:bathy_comp}  Bathymetry (in meters) near the Aransas Pass.}
\end{figure}

We develop two distinct finite element meshes (i.e., domain discretizations) to use in the ADCIRC model to analyze the effects of a change in the ship channel’s bathymetry on hydrodynamics in the domain. One mesh utilizes the current channel bathymetry, and the other uses the proposed bathymetry. Aside from the variation in bathymetry, the two meshes are identical in terms of span and resolution. The current and proposed bathymetries in the Aransas Pass are shown in Figure~\ref{fig:bathy_comp}.
In order to ensure accurate results, these meshes span the entirety of the Gulf and the western Atlantic Ocean with particularly high resolution along the Texas coast. The ADCIRC model used with the current bathymetry mesh has been extensively validated and is frequently employed in operational storm surge forecasting, see, e.g.,~\cite{hope2013hindcast}. We also refer to Section 2.2 of this paper for an in-depth discussion on the development of this mesh.
This mesh is shown in Figure~\ref{fig:adcirc_mesh}. Both of the meshes used in this study are constructed specifically for modeling the Texas coast and contain 3,352,598 nodes spread across 6,675,517 finite elements. Resolution along the Texas coast ranges from $100m$ down to $30m$, while the resolution in the deep ocean is of the order of several kilometers. In addition to the offshore and nearshore areas of the coastal ocean, the mesh extends inland to where the land elevation is approximately 10$m$ above NAVD88. The mes also has several of the man made coastal protection systems on the Texas coast, including jetties, seawalls, and levees,  and uses a weir formula to assess overtopping of these, see~\cite{luettich1995assessment}. Finally, to capture the inundation of dry land and flow over land, ADCIRC has an advanced algorithm for wetting and drying of the finite elements~\cite{dietrich2004assessment}.

With the large domain and significant number of nodes and finite elements, ADCIRC computations relies on high-performance computing clusters at the Texas Advanced Computing Center (TACC). ADCIRC is parallelized based on domain decomposition and the MPI (Message Passing Interface) library. Details on the parallelization and scaling properties can be found in e.g., ~\cite{Tanaka:2010,Dietrich:2012b} and we also present a scaling study in Section~\ref{sec:scaling} below. 

\subsection{ADCIRC Model Scaling Properties} \label{sec:scaling}

To illustrate the parallel performance of ADCIRC, we perform a study of its strong scaling behavior using the Frontera supercomputer at TACC~\cite{stanzione2020frontera}. Frontera consists of Intel 8280 “Cascade Lake” computational nodes with 56 processors per node. 
For this study, we use the same computational mesh used in this study,  described further in Section\ref{sec:modeling_calval} and shown in Figure~\ref{fig:adcirc_mesh} which consists of 3,352,598 nodes and 6,675,514 triangular elements and perform a simulation of tidal fluctuations over a 12 hour period using a 1$s$ time step.
\begin{figure}[h]
  \centering
  \includegraphics[width=0.65\textwidth]{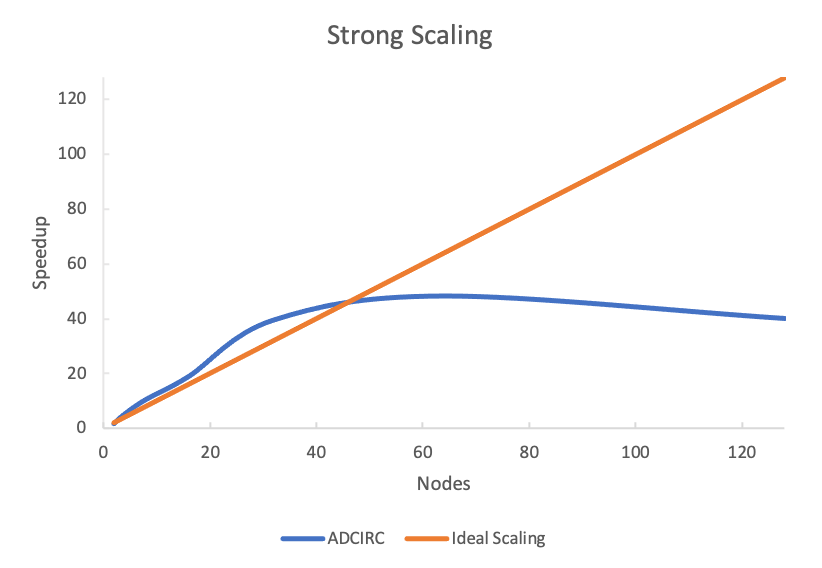}
  \caption{Parallel performance of ADCIRC}
  \label{fig:ADCIRC_SCALING}
\end{figure}
For this particular mesh, the speedup is nearly ideal up to 50 compute nodes as shown in Figure~\ref{fig:ADCIRC_SCALING}. This type of behaviour has also been noted in~\cite{Tanaka:2010} where the speedup becomes sub-optimal once the mesh partition has less than 2000 finite elements per processor. The degradation is due to the increased cost of communication in relation to the computational cost on each MPI process.

\subsection{ADCIRC Model Set-up}

To obtain an accurate prediction of the circulation throughout the domain, the ADCIRC models developed here use multiple external forcings. Critical to the circulation are tides, wind, pressure,  Coriolis force, and bottom friction. These forces appear in the right hand side of the momentum equations~\eqref{eq:SWE}, in $F_x,F_y$. Coriolis force is obtained from based on a classical relation between the force and Latitude, see, e.g., \cite{luettich1992adcirc} is is therefore the most trivial force to ascertain. Bottom friction in the SWE can be based on multiple constitutive relations between water depth, velocity, and classification of the bottom type. In the present work, we use the commonly adopted Manning's $n$ friction formulation, and the land cover classifications are used for spatially varying values for $n$, see~\cite{hope2013hindcast} for details on the data sources.    
Tidal forcing is applied both as a body force  in $F_x,F_y$ in~\eqref{eq:SWE}, and as a boundary condition at the eastern portion of the mesh along the 60$^\circ$ meridian. 
To ascertain these forcings, we use a software package developed specifically for ADCIRC meshes called OceanMesh2D~\cite{Pringle2018}, which employs the TPXO9 tidal model~\cite{Egbert2002} to provide the tidal forcing for the correct period of time. Lastly, winds at 10 meters above the water surface and air pressure are obtained from The North American Mesoscale Model (NAM) at  six hour increments. After obtaining NAM data from \url{https://www.ncei.noaa.gov/data/north-american-mesoscale-model}, the data is then transformed into an ADCIRC readable format and subsequently interpolated  onto the finite element mesh. 

After a validation of this model set-up for the current channel bathymetry, we select   three different time periods to simulate and study as circulation patterns are significantly impacted by large-scale, global meteorological trends. We have selected December 2020, June 2021, and August 2021 to represent three variations in hydrodynamics due to seasonality. Additionally, we considered two extreme cases: Hurricane Nicholas, a Category 1 hurricane which made landfall in mid-September 2021~\cite{Latto:2022}, as well as Hurricane Harvey, a Category 4 hurricane which made landfall mid-August 2017~\cite{goff2019outflow}. For each of these time frames, we use both aforementioned meshes in our ADCIRC models to establish the hydrodynamics for each channel bathymetry.

\subsection{ADCIRC Model Calibration and Validation}
\label{sec:modeling_calval}

\begin{figure}[h!]
\centering
 \includegraphics[width=0.75\textwidth]{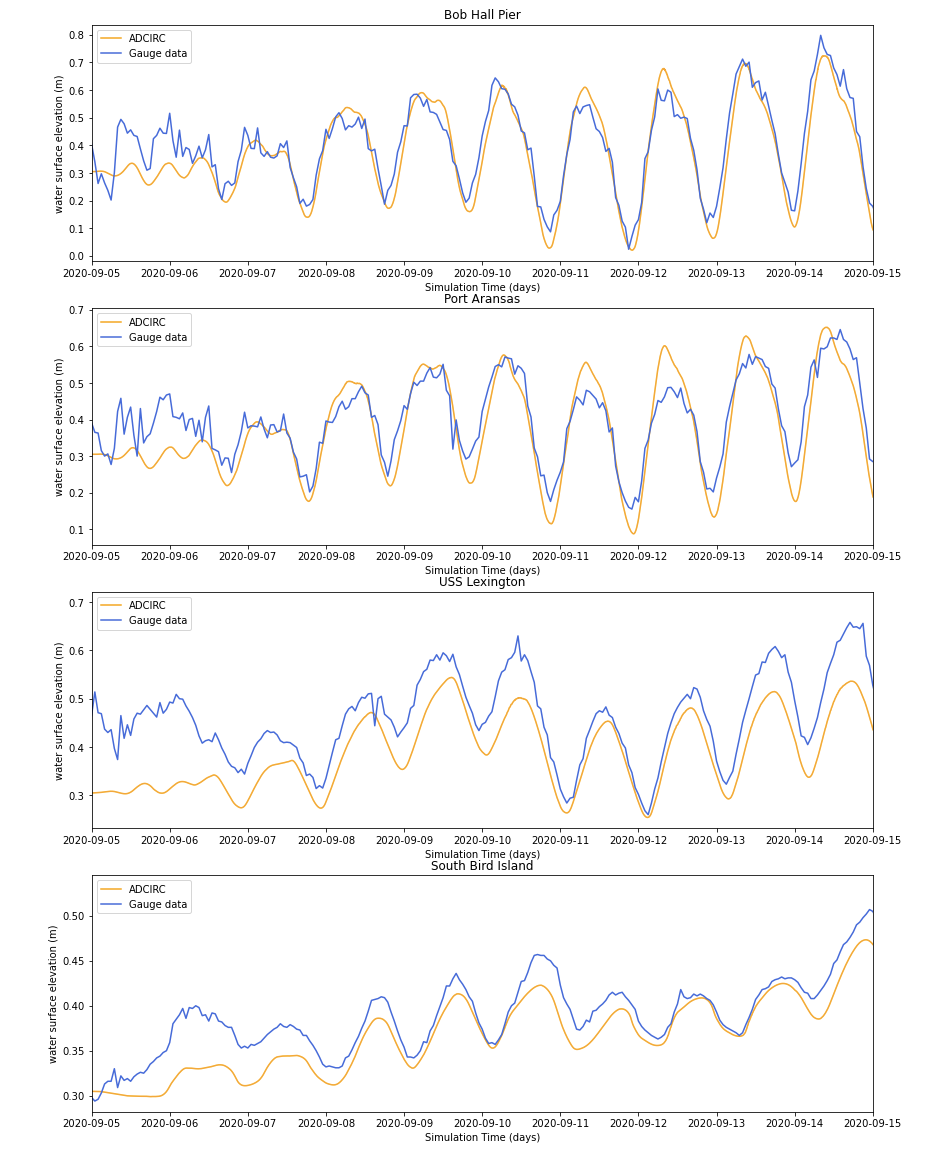}
  \caption{\label{fig:adcirc-elev} Water surface elevations in ADCIRC compared to nearby NOAA gauges during September 2020 for validation }
\end{figure}

To calibrate the ADCIRC model and ensure its physical correctness with the tidal and meteorological forcings, we consider the mesh developed for the current channel bathymetry and compare the elevation output with water elevation data from four NOAA gauges along the Texas coast: Port Aransas (ID 8775237), USS Lexington (ID 8775296), Bob Hall Pier (ID 8775870), and South Bird Island (ID 8776139). The model is validated over a 10-day period from September 5, 2020 to September 15, 2020. In Figure~\ref{fig:adcirc-elev}, we show a comparison between the simulated elevation data and observations from the NOAA gauges.
Note that the initial discrepancy  during the first two days is mainly due to a ramping up of the tidal forcing that is necessary to preserve numerical stability. After this ramp period, the simulations agree with the measurements to a level we consider satisfactory for the present study.

\subsection{Particle Transport}
\label{sec:modeling_particles}

After the ADCIRC model is used to simulate the velocity distribution in our domain, a Lagrangian particle tracking code is used to establish the trajectory of oil particles. This particular code is developed and described in~\cite{dietrich2012surface,cyriac2020wind}, and it considers the convective transport of particles through the domain. 

\begin{figure}[h!]
\subfigure[ \label{fig:particles_onshore} Initial condition I - 42,000 particles initially located at Harbor Island.]{\centering
\includegraphics[width=0.4\textwidth]{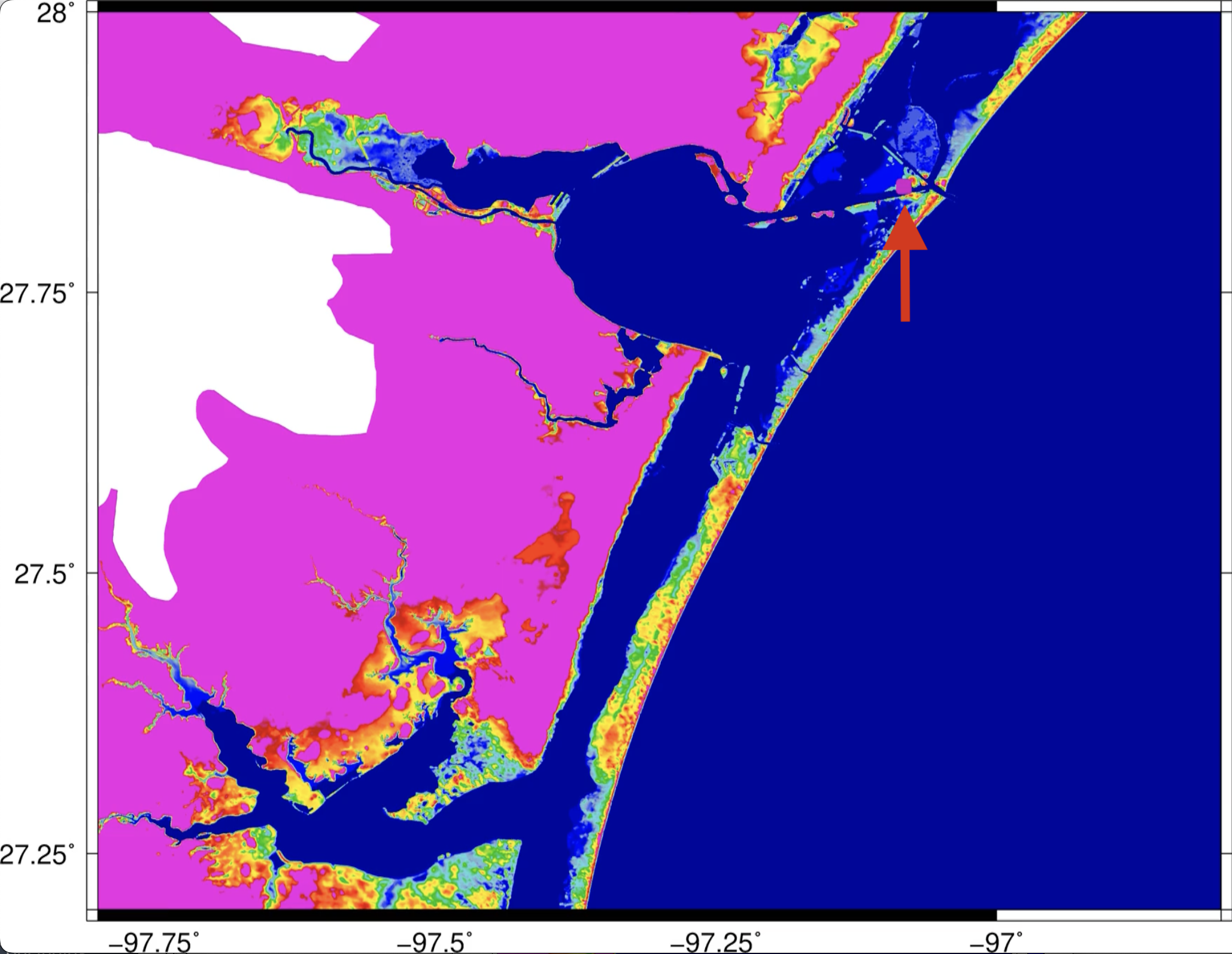}} \hspace*{0.4cm}
\subfigure[ \label{fig:particles_offshore_v1}  Initial condition II - 42,000 particles initially located 21 miles east of the Port of Corpus Christi. ]{\centering
\includegraphics[width=0.48\textwidth]{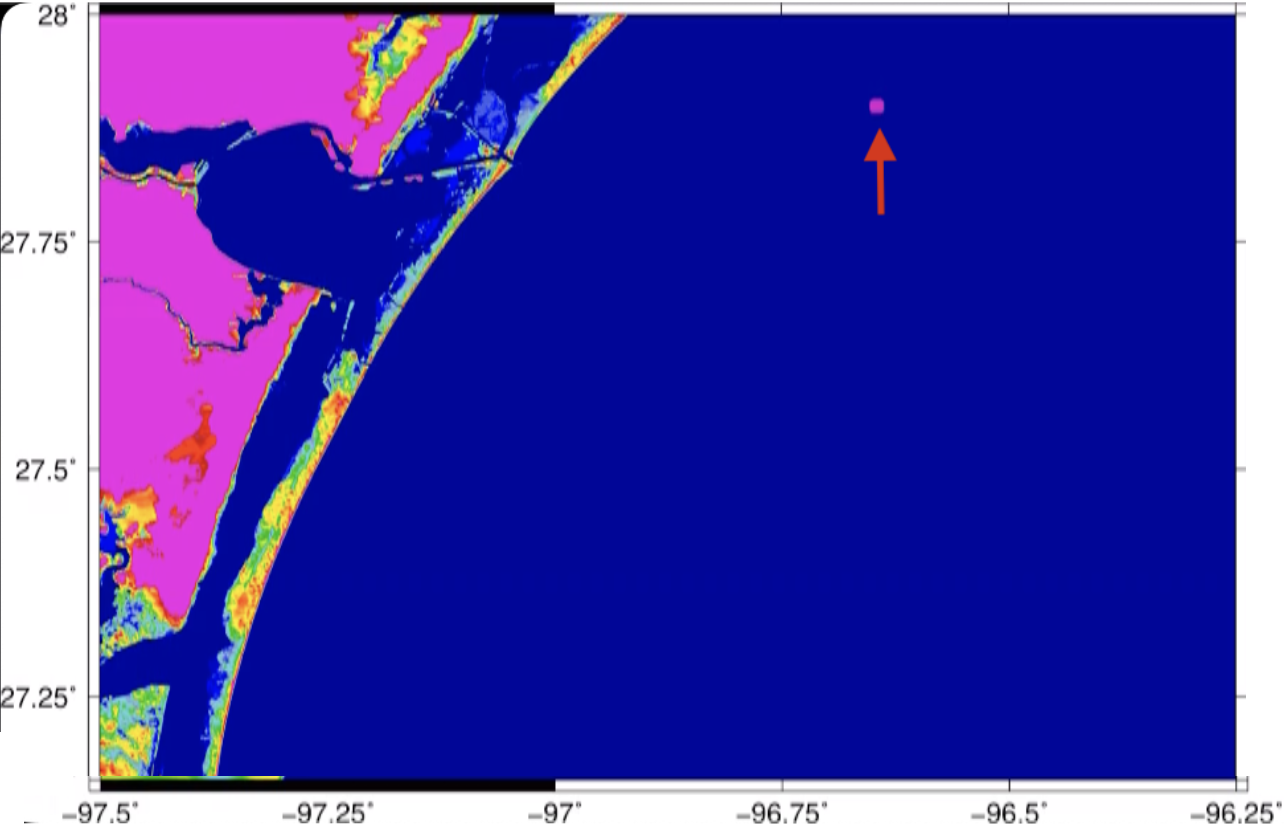}}
\caption{\label{fig:particles_offshore} Initial particle locations. The color spectrum denotes bathymetry above NAVD88 in meter.}
\end{figure}

This particle tracking code requires two inputs: the velocity fields from ADCIRC and the initial distribution of particles in the domain. Initial particle locations are based on the proposed loading sites and are shown in Figure~\ref{fig:particles_offshore} (see Figure~\ref{fig:terminal_locations} for comparison). 
The total number of particles represents the total number of gallons of oil in a theoretical spill, i.e., one particle per gallon. The pipeline used to fill tankers at both locations pump approximately 28,000 gallons per minute. Since an automatic shut-off valve would detect a failure and shut off upstream flow in 1-2 minutes, we therefore assume a 42,000-gallon spill and 42,000 particles to be released at the start of the simulation. 
%
%

%
% VLCC (onshore): 27.844388, -97.082263
% BWTX (offshore): 27.899092, -96.644778
%

% IF WE END UP USING THE LOCATE PARTICLES CODE:
% In Figure~\ref{fig:monitor},  we show the four locations in which we monitor the accumulation of particles  
%
%\begin{figure}[h!]
%\centering
%\includegraphics[width=0.7\textwidth]%{figures/Particle_tracking_highlight.jpg}
%\caption{\label{fig:monitor} Monitoring locations, color spectrum denotes bathymetry above NAVD88 in meter. }
%\end{figure}
%
%
%To analyze the data from the particle tracking algorithm, we developed a tool that monitors the particle location throughout the simulation. Once a particle reaches a suitable habitat, its velocity is set to zero and we consider it to be a successfully recruited larvae. This is a rather simplistic approach, however, our goal here is to ascertain if the number of larvae that reach the seagrass beds is impacted by the changed channel bathymetry. 
% 

\section{Results}
\label{sec:results}

Here, we present results of the particle trajectories in the regions near the Port of Corpus Christi. First, we consider the trajectories of particles released at the two loading sites in the months of December 2020, June 2021, and August 2021. Both the current and proposed channel depths are considered for each time period. We then consider particle trajectories under extreme flow conditions, particularly the hydrodynamic conditions during Hurricane Nicholas and Hurricane Harvey.

\subsection{Normal Flow Conditions}
\label{sec:res_particle}

% 1. channel depth: \\
% - no noticeable difference between existing and proposed channel depths on particle trajectories\\
% - particles injected at the offshore site stay in the Gulf of Mexico, with the exception of the Dec 2020 cases (particles begin to cross the barrier island after 22 days)\\
% - particles injected at the onshore loading location spread along the national seashore\\
%2. seasonality:\\
%- winter (Dec 2020) onshore - slightly more spread throughout the national seashore compared to summer \\
% 3. extreme weather events\\
% - category 1 - > large spread from onshore site throughout National Seashore, but offshore stay in gulf \\

% In Summary:\\
% - spread of particles depends on onshore vs offshore (can we call these ICs?) and hydrodynamic conditions (seasonality, extreme weather conditions, not really channel depth)\\
% \textbf{if we decide to use the particle tracking code, see section in larvae report}\\

\begin{figure}[h!]
\subfigure[ \label{fig:existing_dec_onshore_end} Current bathymetry and onshore site.]{\centering
\includegraphics[width=0.47\textwidth]{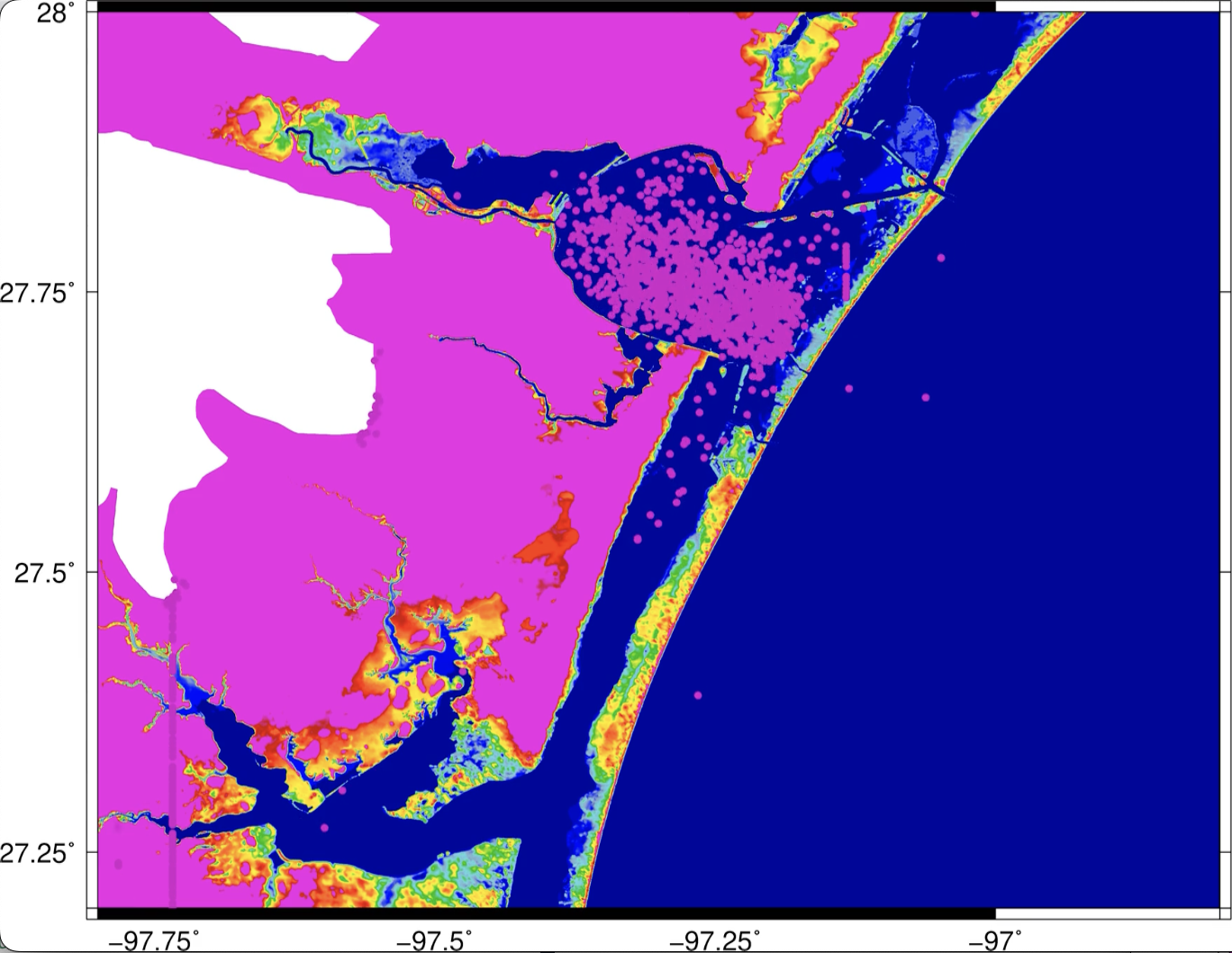}} \hfill
\subfigure[ \label{fig:future_dec_onshore_end}  Proposed bathymetry and onshore site. ]{\centering
\includegraphics[width=0.42\textwidth]{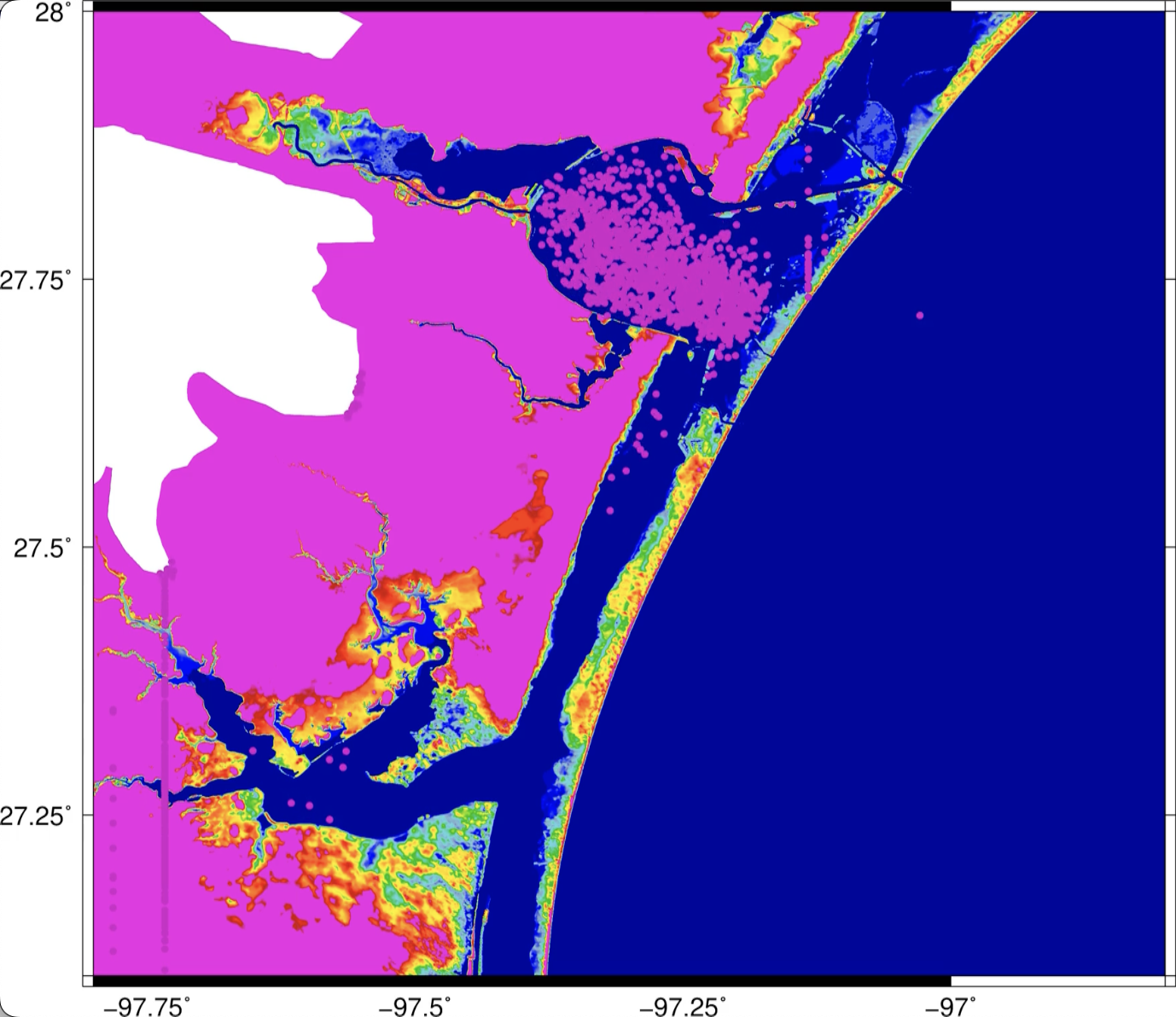}}
\subfigure[ \label{fig:existing_dec_offshore_end_v1} Current bathymetry and offshore site]{\centering
\includegraphics[width=0.43\textwidth]{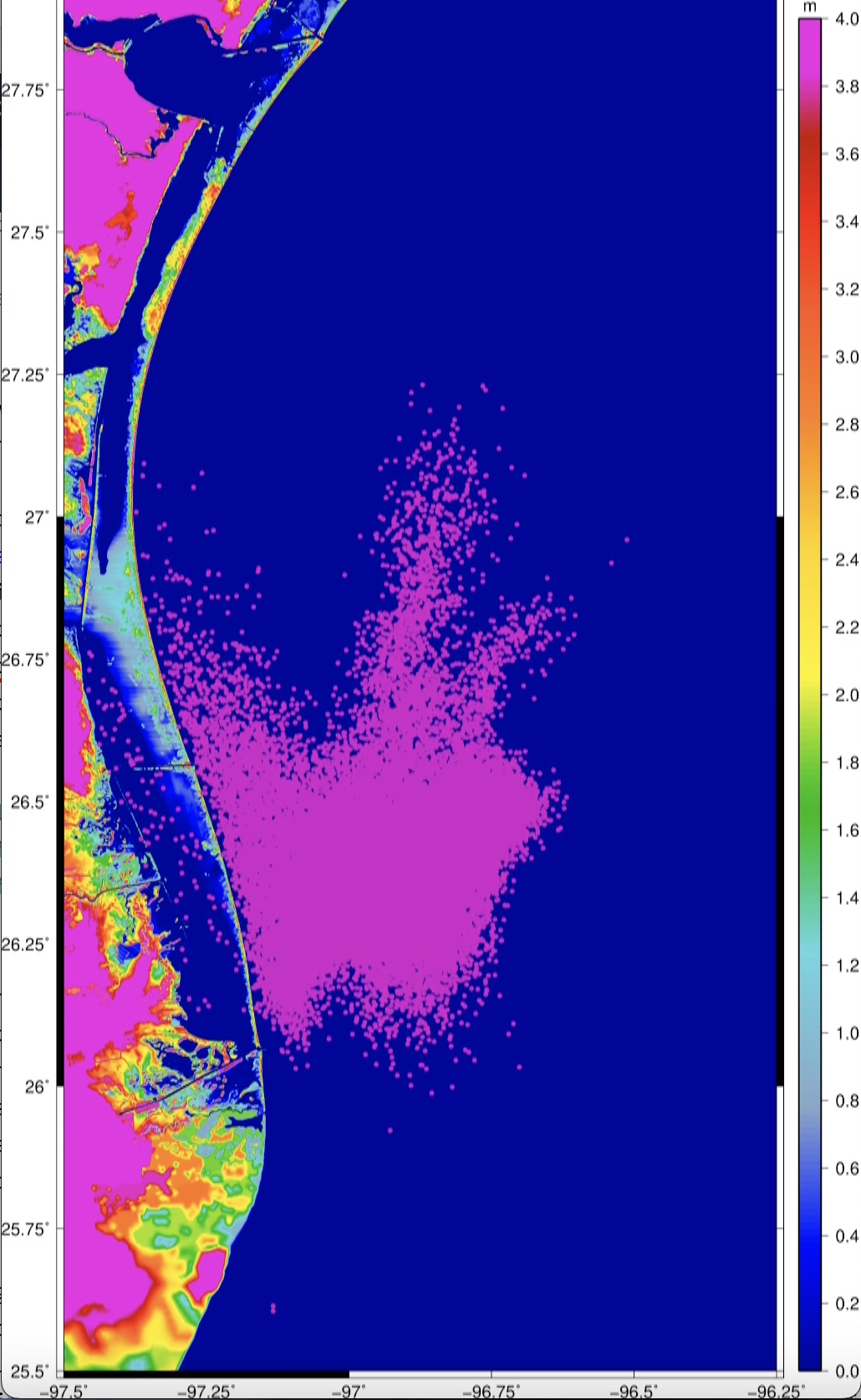}}\hfill
\subfigure[ \label{fig:future_dec_offshore_end_v2} Proposed bathymetry and offshore site]{\centering
\includegraphics[width=0.43\textwidth]{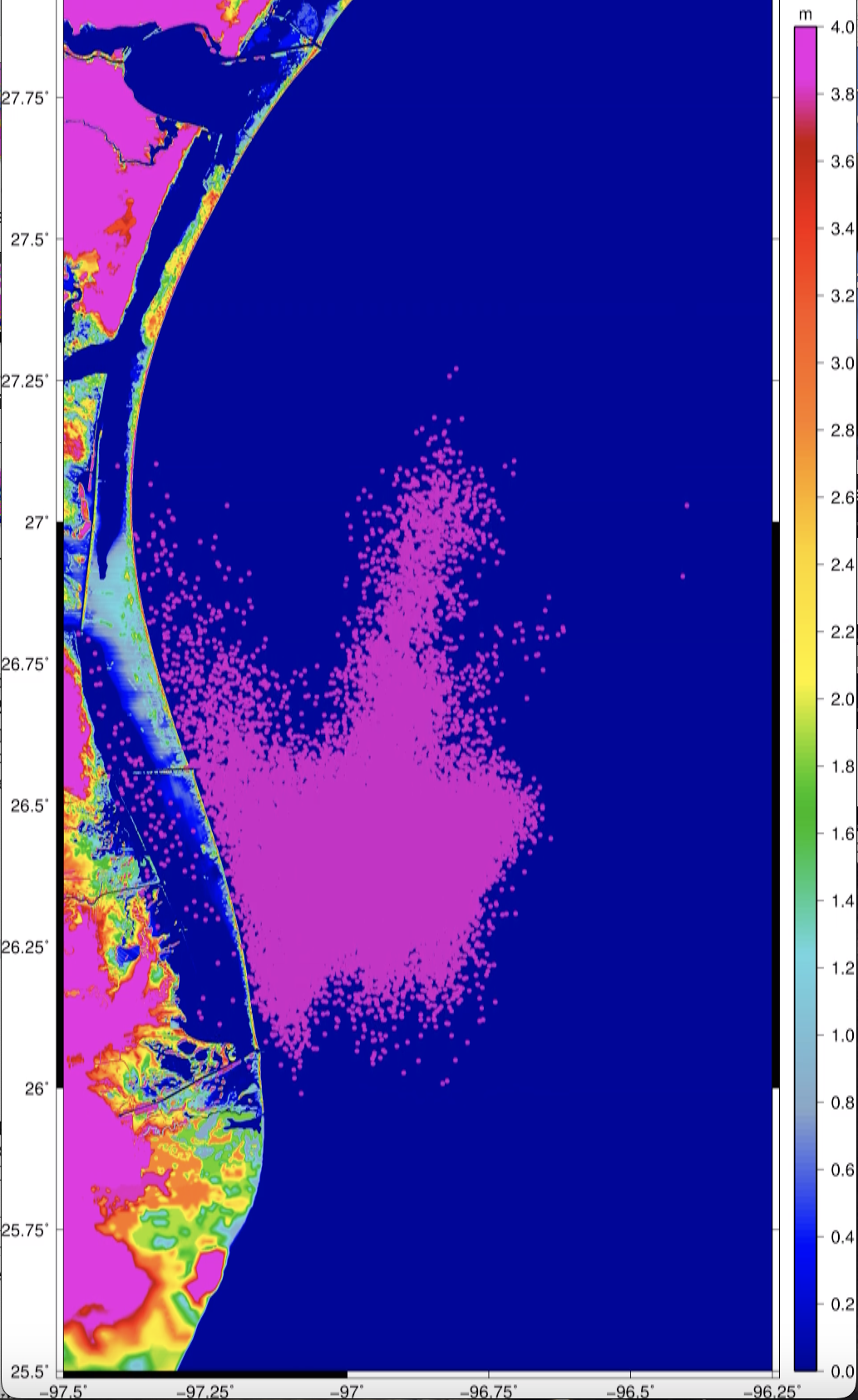}}
\caption{\label{fig:bathymetry_comparison} Particle locations on December 31, 2020, thirty days after the initial release of particles. The color spectrum denotes bathymetry above NAVD88 in meter.}
\end{figure}

\begin{figure}[H]
\subfigure[ \label{fig:existing_dec_offshore_end} Particle locations at the of end of December 2020.]{\centering
\includegraphics[width=0.3\textwidth]{figures/existing_dec_offshore_end.png}}
\hspace*{0.25cm}
\subfigure[ \label{fig:existing_june_offshore_end} Particle locations at the end of June 2021. ]{\centering
\includegraphics[width=0.3\textwidth]{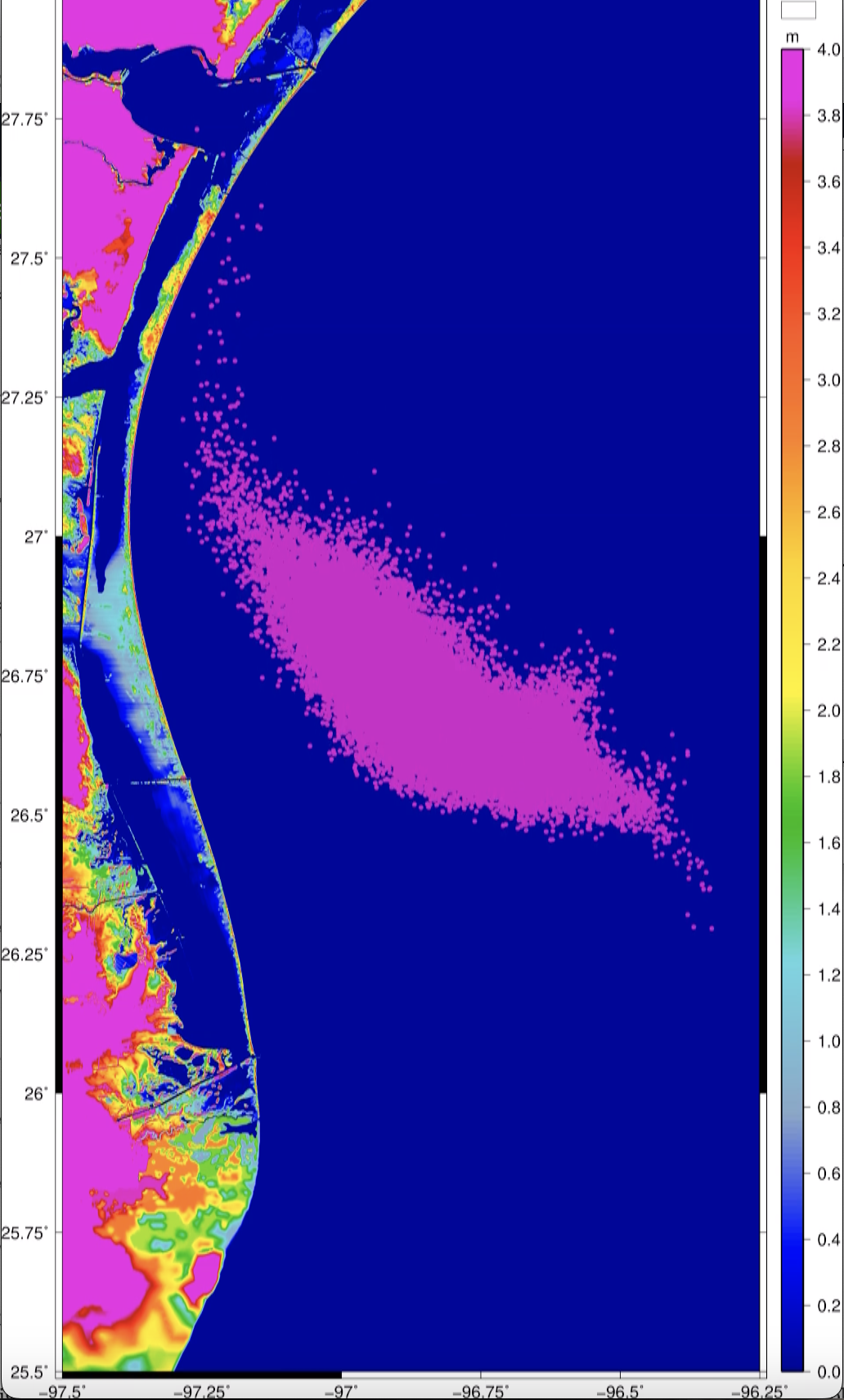}}
\hspace*{0.25cm}
\subfigure[ \label{fig:existing_aug_offshore_end} Particle locations at the end of August 2021.]{\centering
\includegraphics[width=0.3\textwidth]{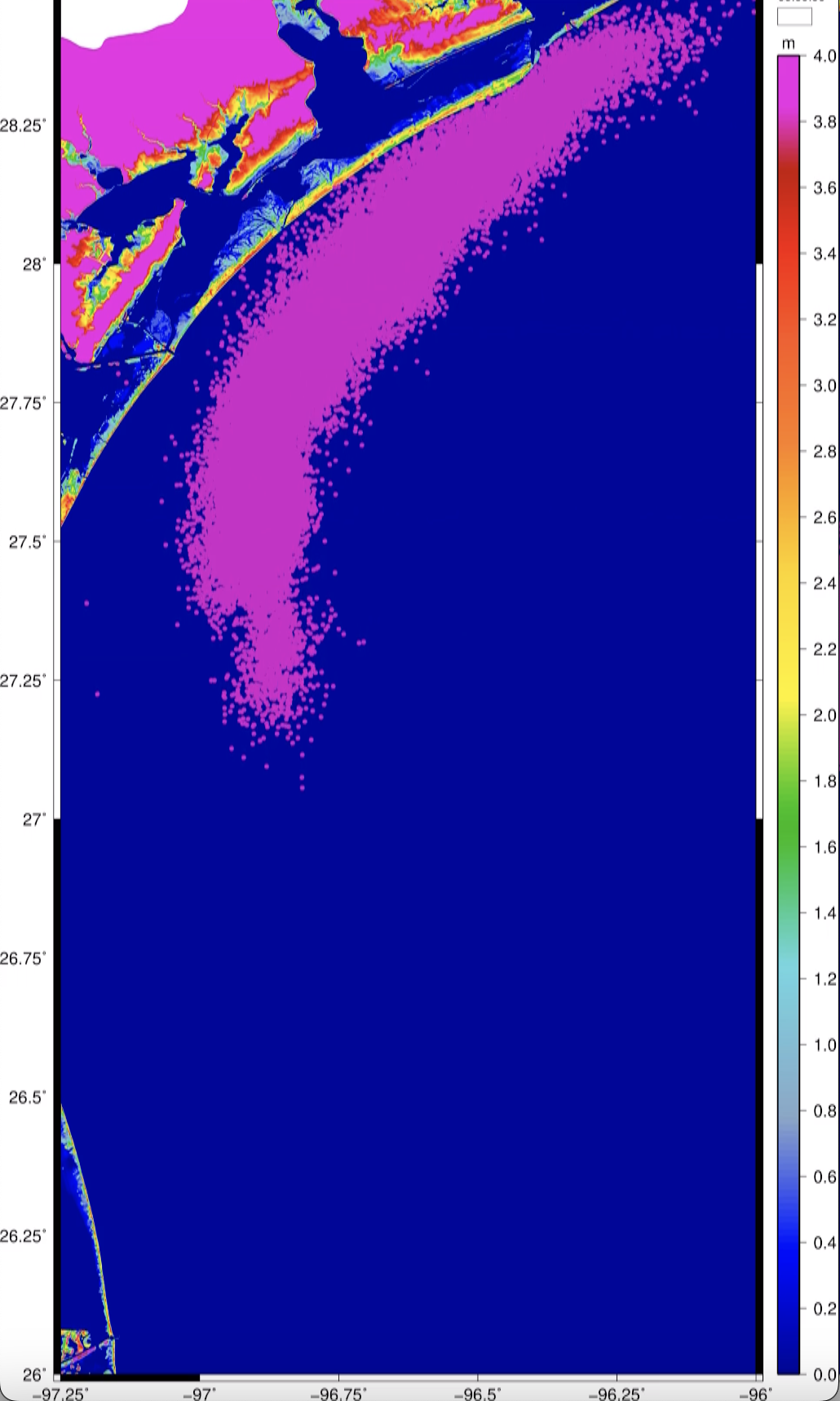}}
\caption{\label{fig:offshore_comparison} Particle locations at the end of runs for each time period using the current bathymetry and an initial location at the offshore loading site. The color spectrum denotes bathymetry above NAVD88 in meter.}
\end{figure}

\begin{figure}[H]
\subfigure[ \label{fig:future_dec_onshore_5days} Particle locations on December 6, 2020.]{\centering
\includegraphics[width=0.45\textwidth]{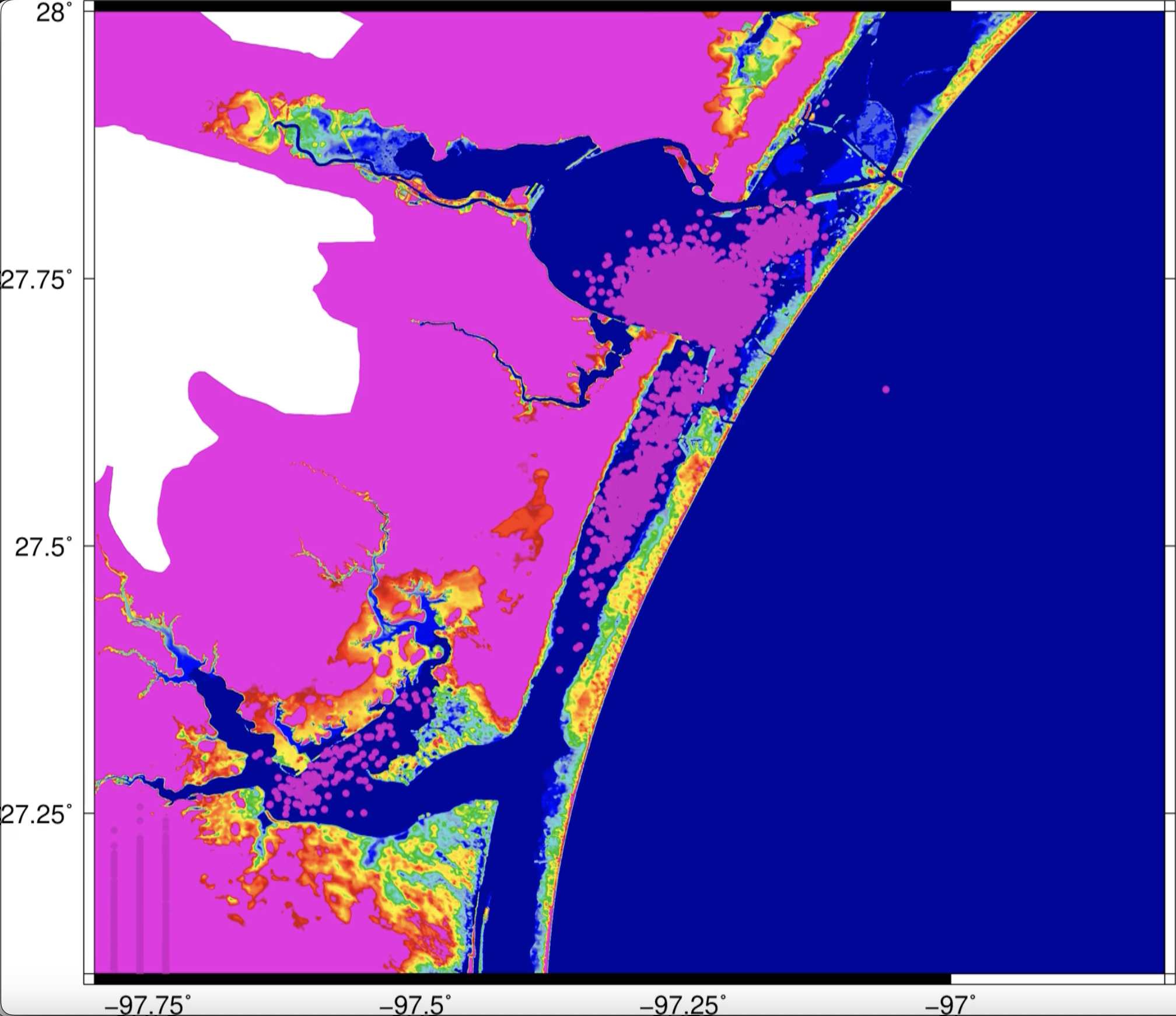}}
\subfigure[ \label{fig:future_june_onshore_5days} Particle locations on June 6, 2021. ]{\centering
\includegraphics[width=0.45\textwidth]{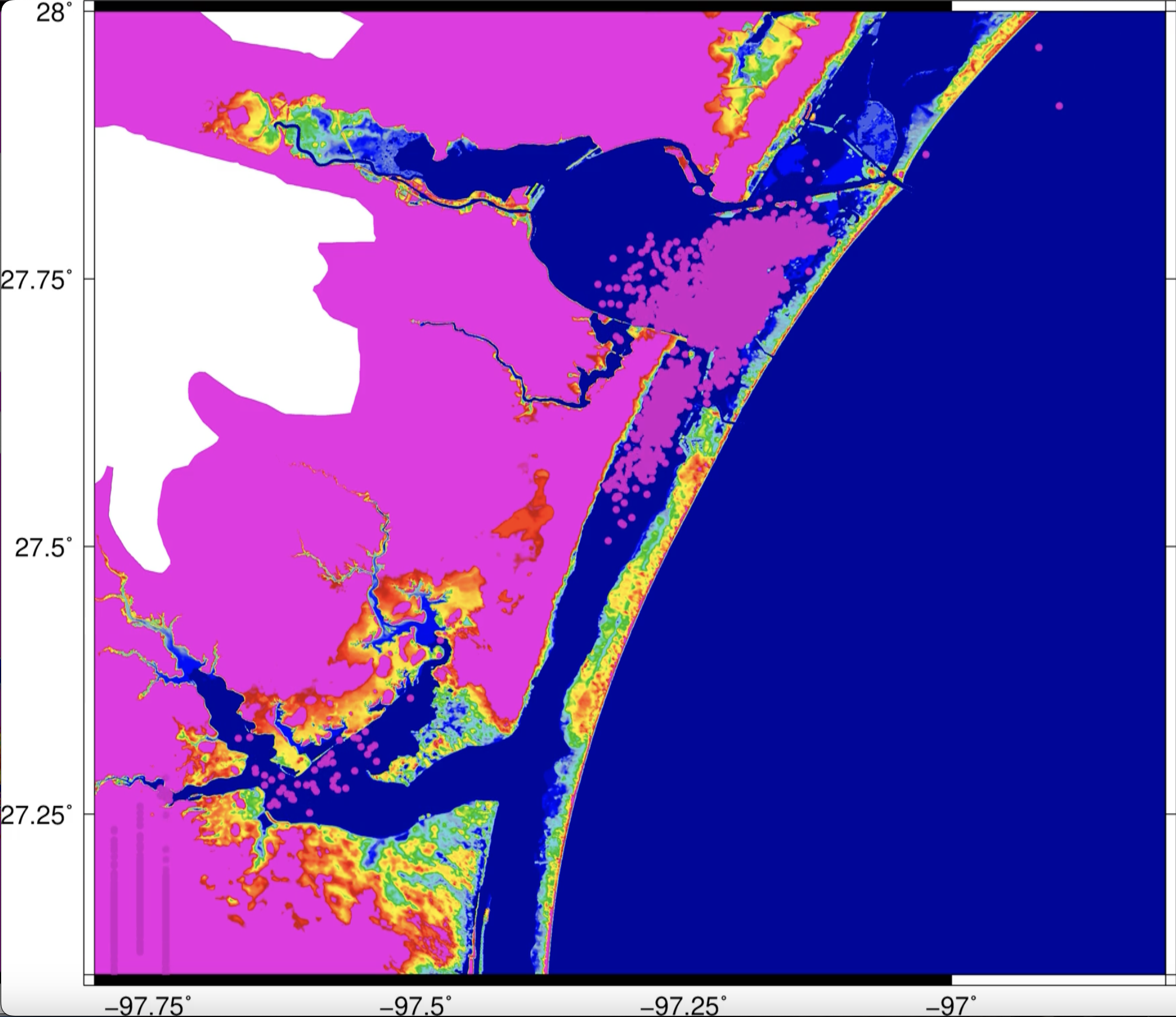}}
\subfigure[ \label{fig:future_aug_onshore_5days} Particle locations on August 6, 2021.]{\centering
\includegraphics[width=0.45\textwidth]{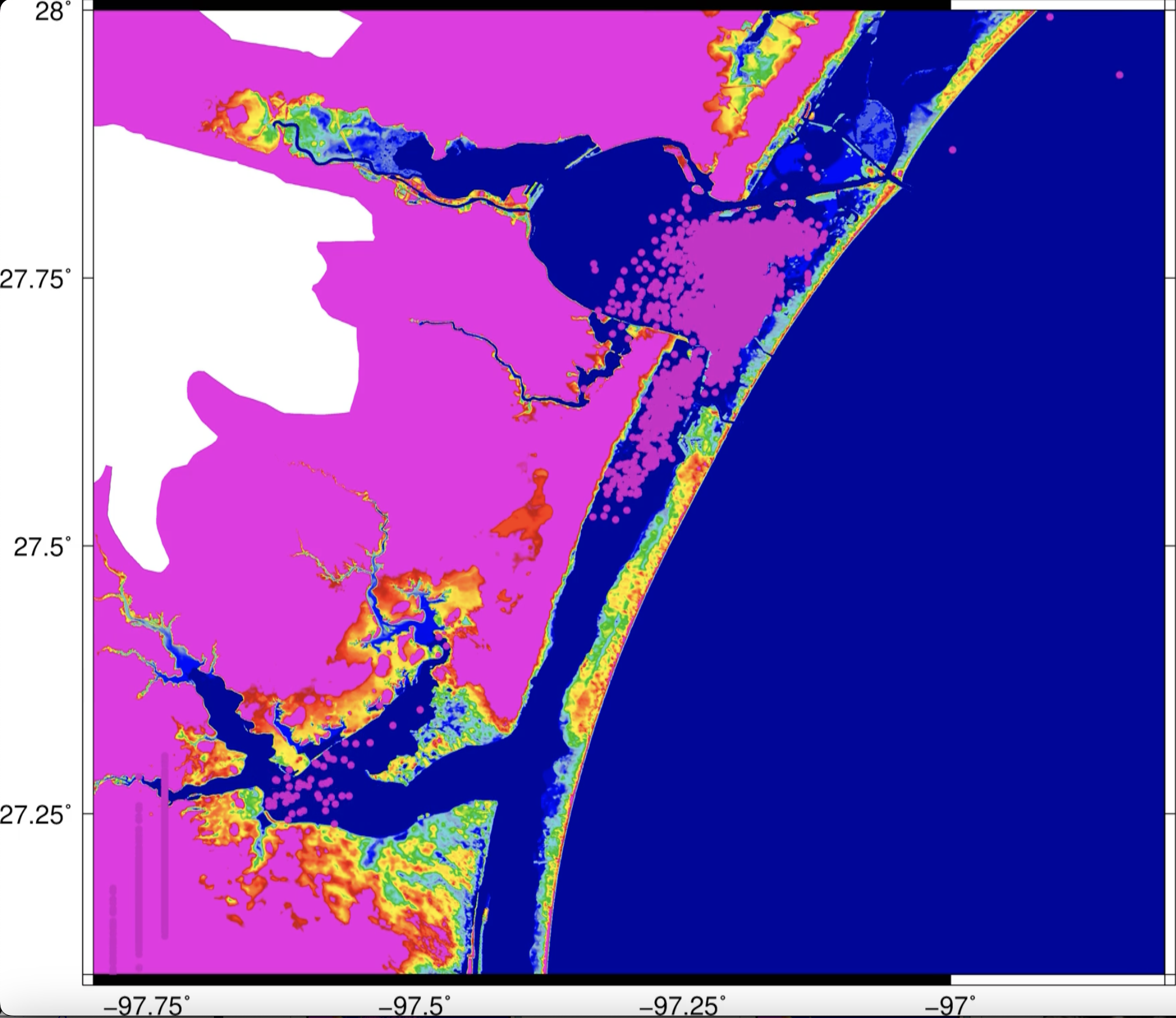}}
\caption{\label{fig:onshore_comparison} Particle locations six days after model start time for each time period using the proposed bathymetry and an initial location at the onshore loading site. The color spectrum denotes bathymetry above NAVD88 in meter.}
\end{figure}

For each considered time period (December 2020, June 2021, and August 2021) and for both bathymetries, we assess the distribution of particles after 30 days in a qualitative manner. In particular, we compare the visual difference when inspecting the outputs. Overall, the  results suggest that, independent of loading terminal location, a change in bathymetry does not have a significant effect on the particle trajectories, as seen in Figure~\ref{fig:bathymetry_comparison}, which shows December 2020 as an example. 
In general, the number of particles that reach the bays to the west of the barrier islands, with critical coastal ecosystems, is highly dependent on the chosen loading location. Particles injected at the onshore site spread along the Padre Island National Seashore (PINS), and particles injected at the offshore site remain in the offshore region, with the exception of the December 2020 case. During this time, particles begin to cross the barrier island through the channel at Port Mansfield, Texas  after 22 days (see Figure~\ref{fig:offshore_comparison}). 
\noindent From Figure~\ref{fig:onshore_comparison}, it is evident that the extent of particle spread throughout the PINS  from the onshore loading location is dependent upon differences in hydrodynamic conditions caused by seasonality.
\noindent During winter, specifically December 2020 as shown in Figure~\ref{fig:future_dec_onshore_5days}, there is slightly more spread throughout the national seashore in comparison to the two summer months that were considered.

\subsection{Extreme Flow Conditions}
\label{sec:extreme_flow}
% - nick: channel depth doesn't have a large impact; offshore stays entirely offshore
% - harvey: actually a huge difference between existing and future for onshore
% - existing harvey onshore vs offshore: at the end of the 10 days, it looks like a similar-ish number of particles in the bays, but totally different trajectories throughout the 10 days

The overall conclusion of minor impacts of the channel flow in the previous section was drawn based on normal flow conditions where no hurricanes occurred in the area considered. 
However, since the Texas coastline near Corpus Christi is frequently impacted by hurricanes and smaller storms~\cite{frey2010potential}, we consider two cases with extreme flow in the domain. To model such extreme cases, we consider the hydrodynamic conditions during Hurricane Nicholas (2021),  a Category 1 hurricane that made landfall north of Matagorda Bay, and Hurricane Harvey (2017), a Category 4 hurricane as it made landfall north of the ship channel near the city of Rockport, Texas. As these events are isolated extremes, we consider a time span of ten days. While this time span is too short to compare to the 30-day time spans in Section~\ref{sec:res_particle}, it allows us to compare transport from the two proposed loading sites using the two different channel depths during times of extreme flow.

\begin{figure}[h!]
\subfigure[ \label{fig:existing_harvey_onshore_10days} Current bathymetry and onshore loading location]{\centering
\includegraphics[width=0.45\textwidth]{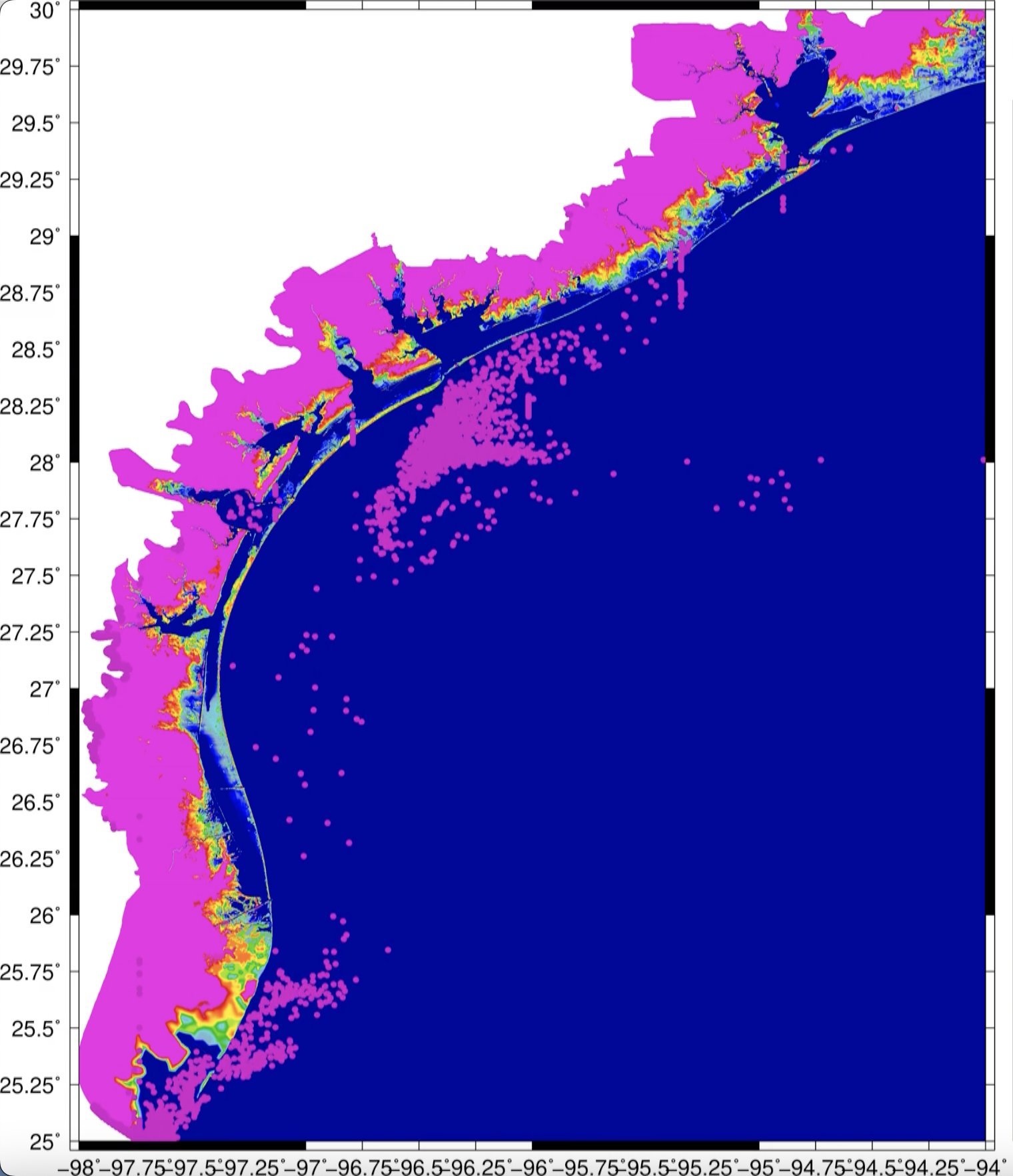}}\hfill
\subfigure[ \label{fig:future_harvey_onshore_10days} Proposed bathymetry and onshore loading location]{\centering
\includegraphics[width=0.45\textwidth]{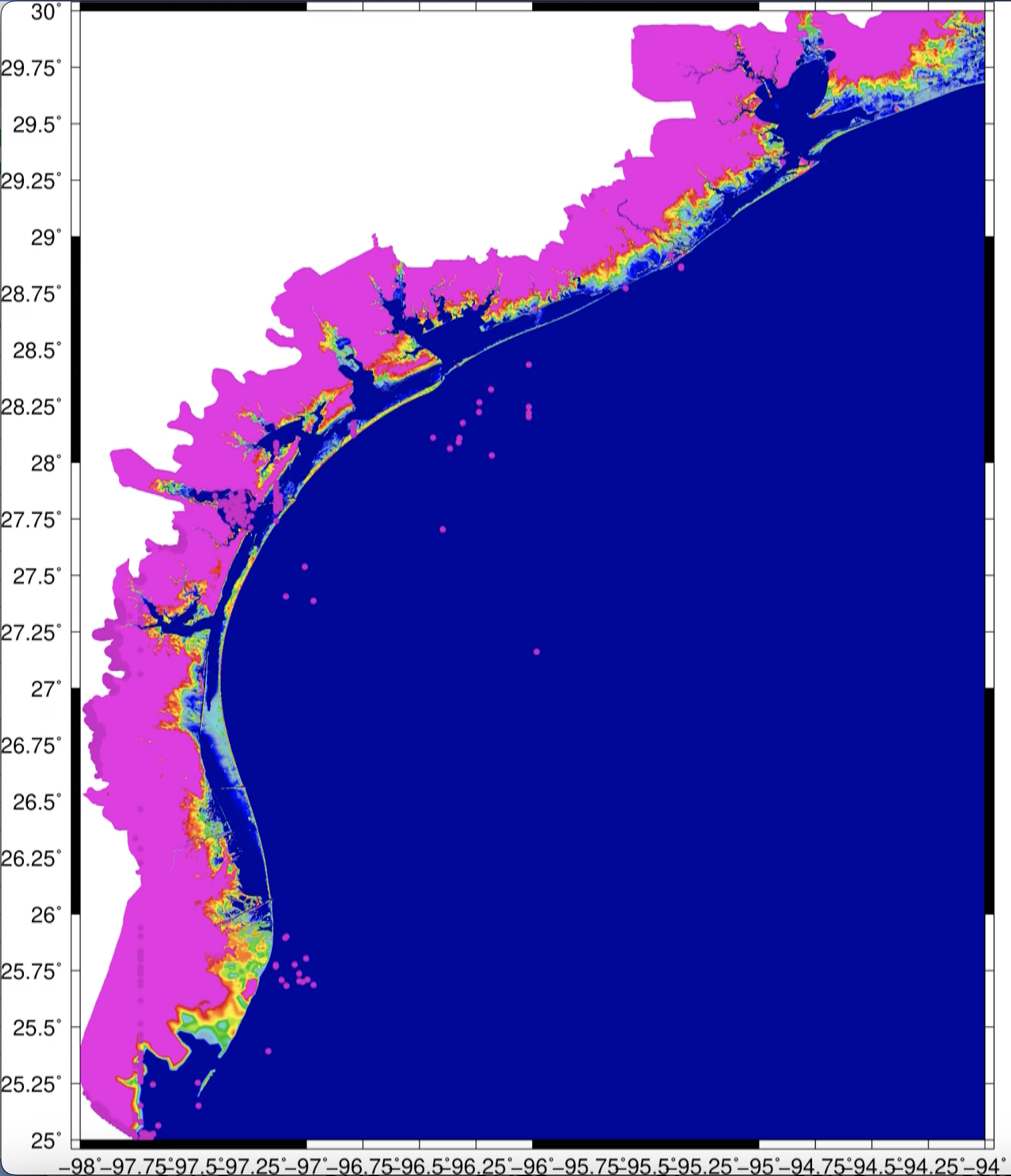}}
\subfigure[ \label{fig:existing_harvey_offshore_10days} Current bathymetry and offshore loading location]{\centering
\includegraphics[width=0.45\textwidth]{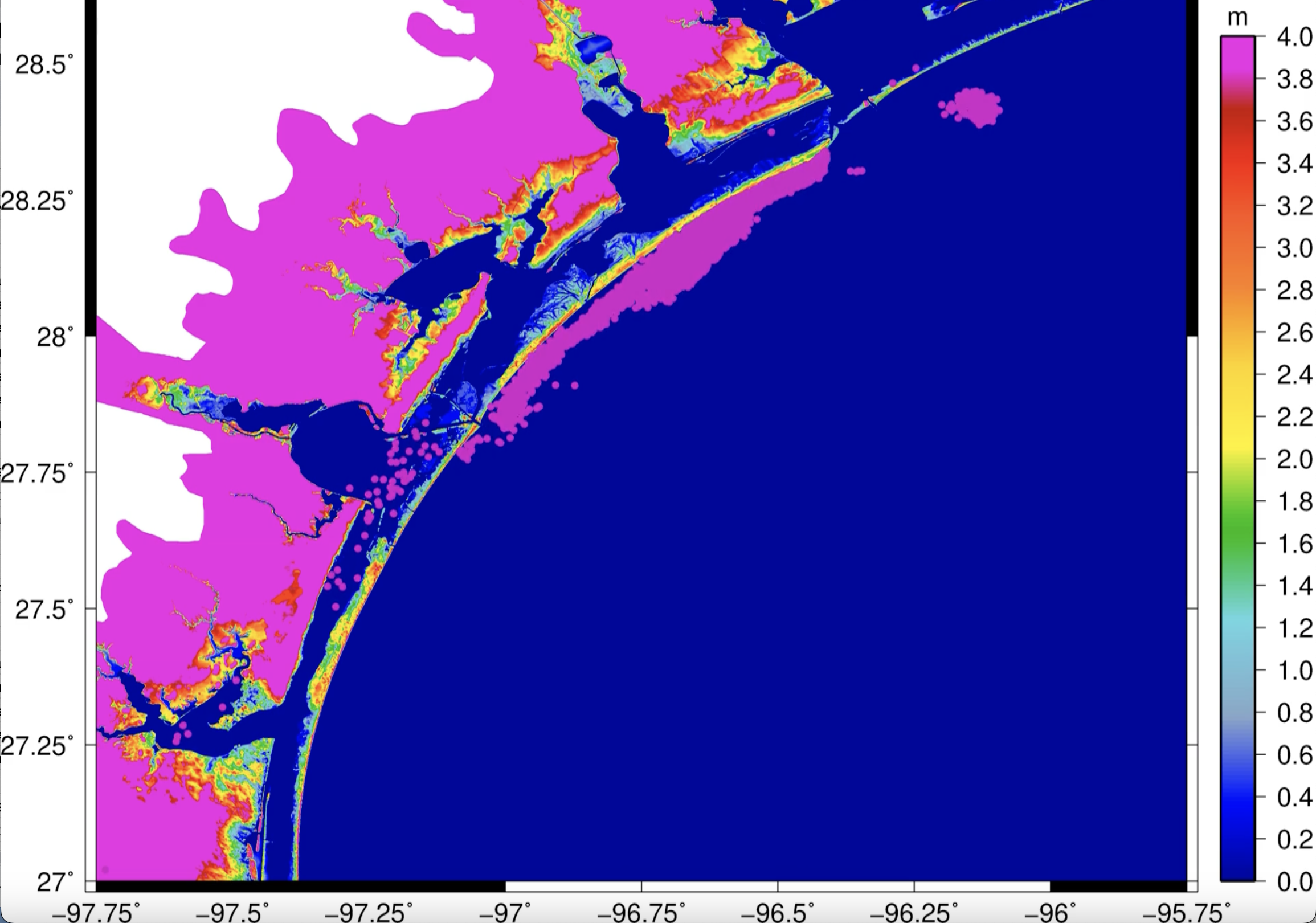}}\hfill
\subfigure[ \label{fig:future_harvey_offshore_10days} Proposed bathymetry and offshore loading location]{\centering
\includegraphics[width=0.45\textwidth]{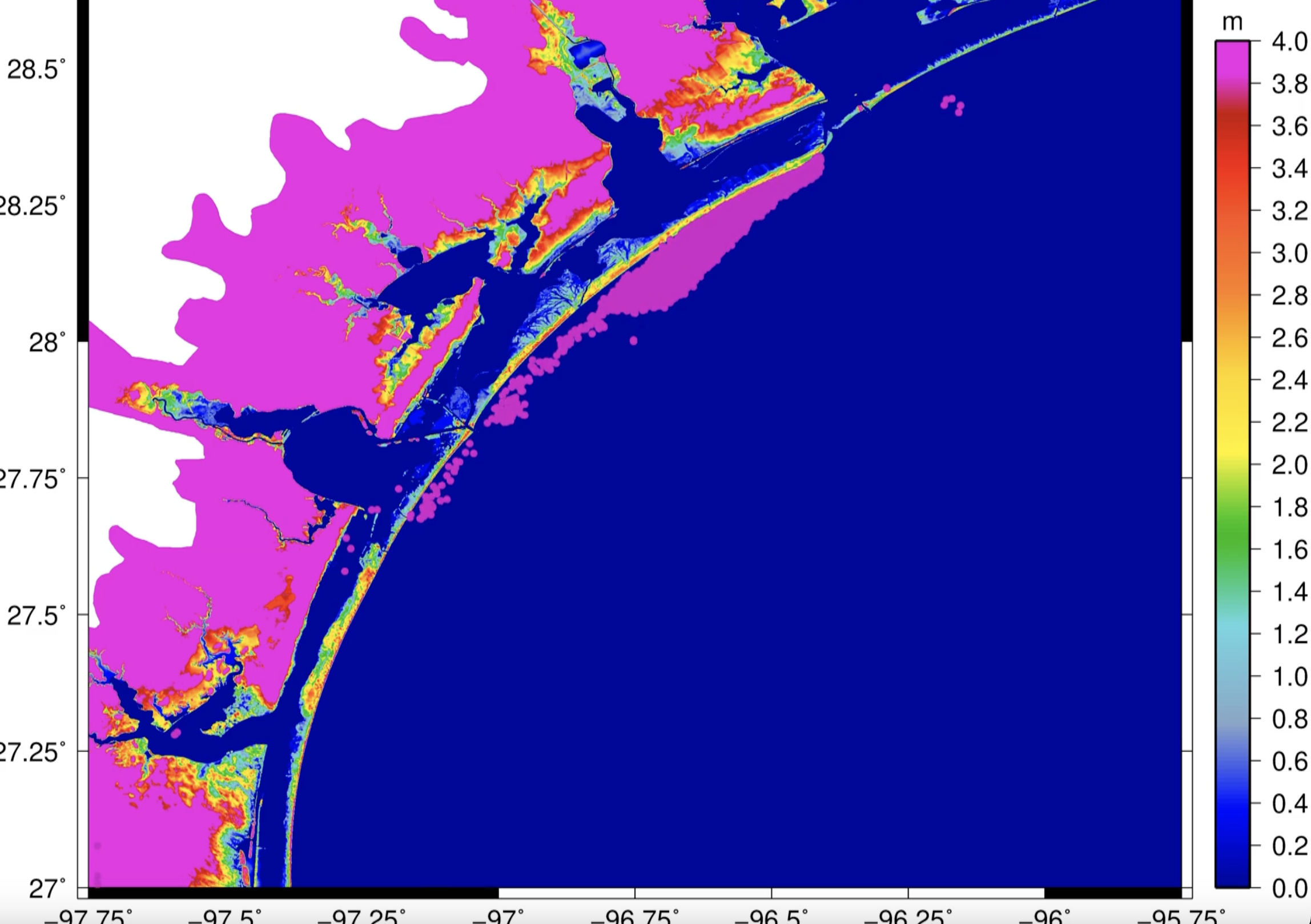}}
\caption{\label{fig:hurricane_harvey_bathymetry} Particle locations ten days after the initial release of particles using the hydrodynamic conditions of Hurricane Harvey. The color spectrum denotes bathymetry above NAVD88 in meter.}
\end{figure}

\begin{figure}[h!]
\centering
 \includegraphics[width=0.75\textwidth]{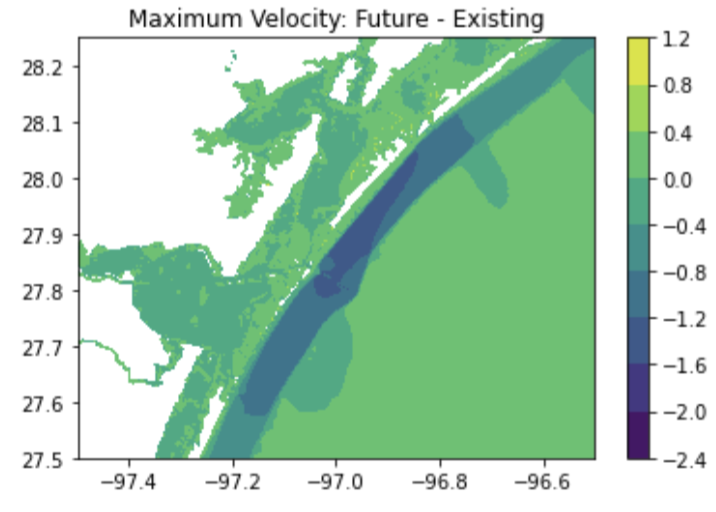}
  \caption{\label{fig:maxvel} The difference in maximum depth-averaged velocities (in m/s) between the two ADCRIC models (i.e., the current and proposed bathymetries) in the region surrounding the Port of Corpus Christi.}
\end{figure}

During Hurricane Nicholas, the particle trajectories follow trends similar to those seen in the "normal" flow cases described in Section~\ref{sec:res_particle}. There is little noticeable difference between the particle trajectories for the current and proposed channel depths. 
\begin{figure}[H]
\subfigure[ \label{fig:existing_nick_onshore_6days} Six days after the initial release of particles from the onshore site using the current bathymetry.]{\centering
\includegraphics[width=0.47\textwidth]{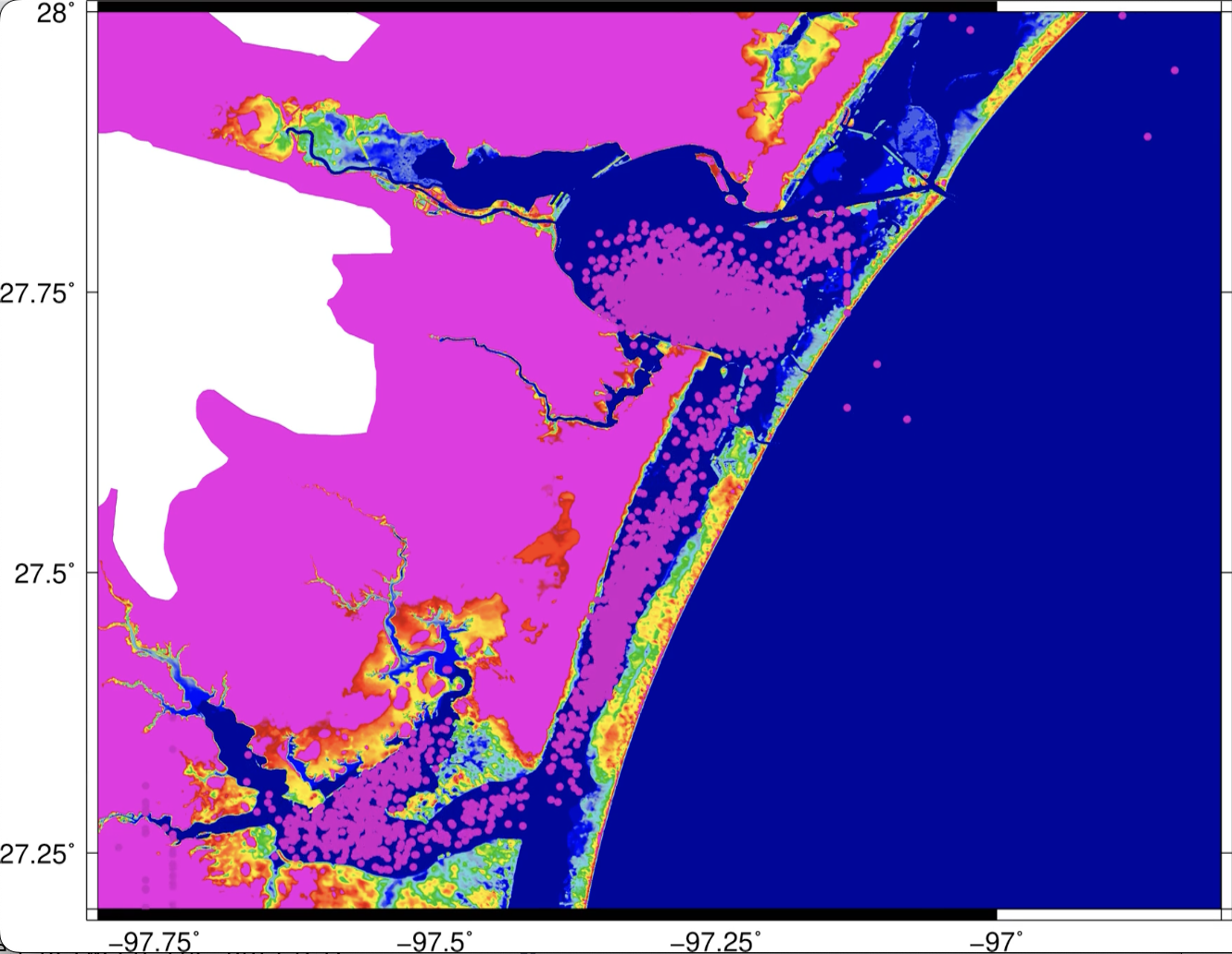}}\hfill
\subfigure[ \label{fig:future_nick_onshore_6days} Six days after the inital release of particles from the onshore site using the proposed bathymetry.]{\centering
\includegraphics[width=0.42\textwidth]{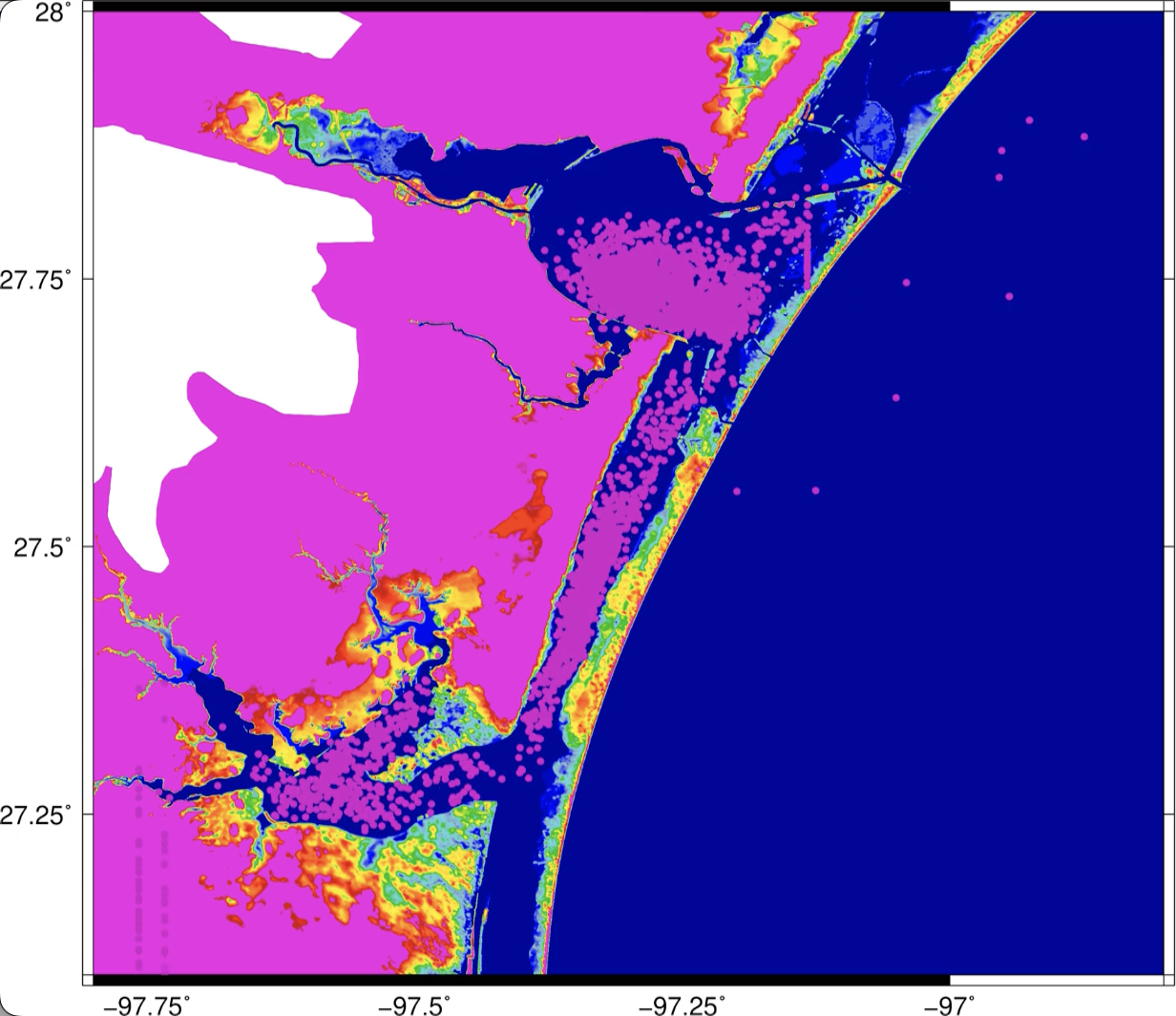}}
\subfigure[ \label{fig:existing_nick_offshore_end} Six days after the inital release of particles from the offshore site using the current bathymetry.]{\centering
\includegraphics[width=0.43\textwidth]{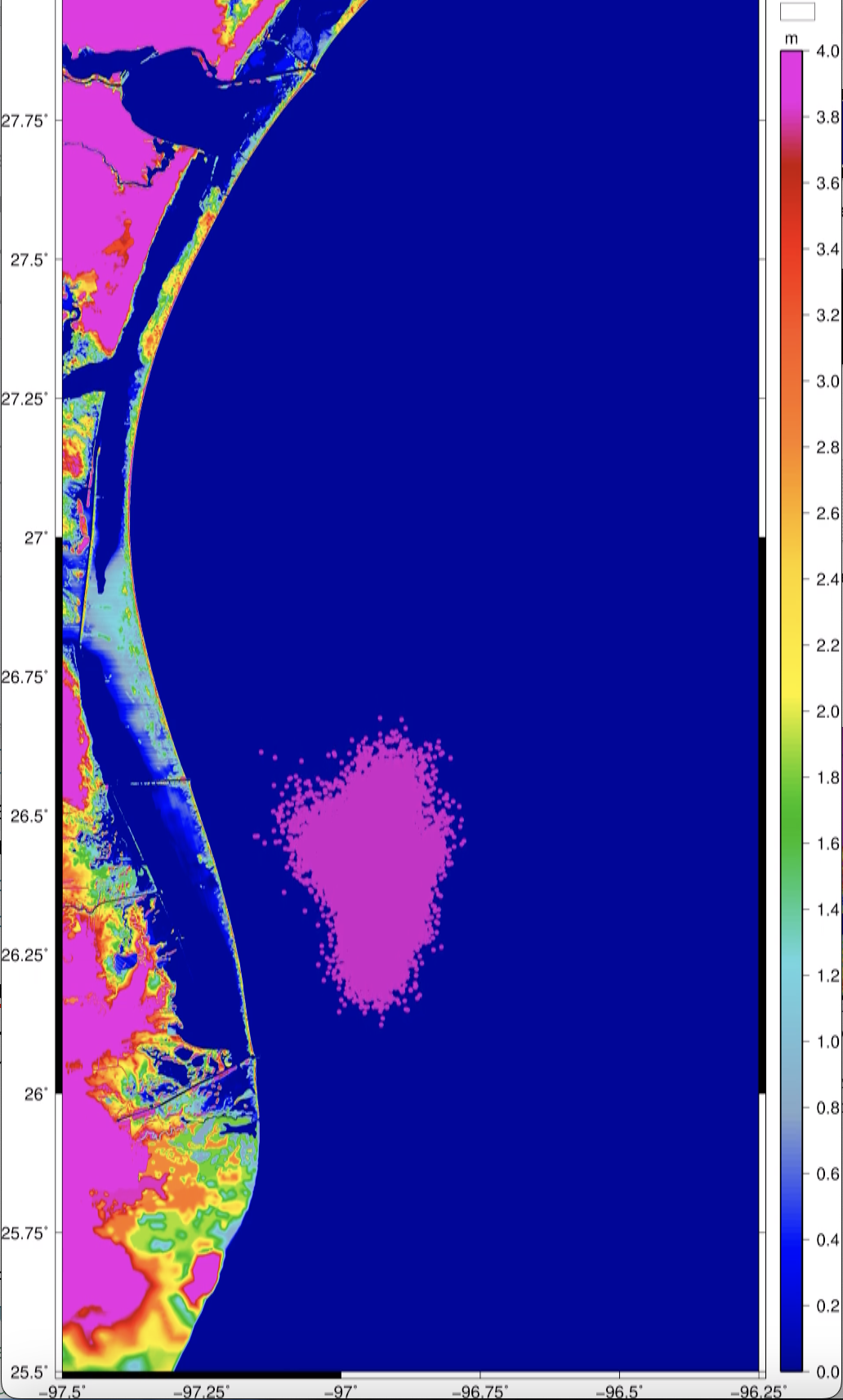}}\hfill
\subfigure[ \label{fig:future_nick_offshore_end} Six days after the inital release of particles from the offshore site using the proposed bathymetry.]{\centering
\includegraphics[width=0.43\textwidth]{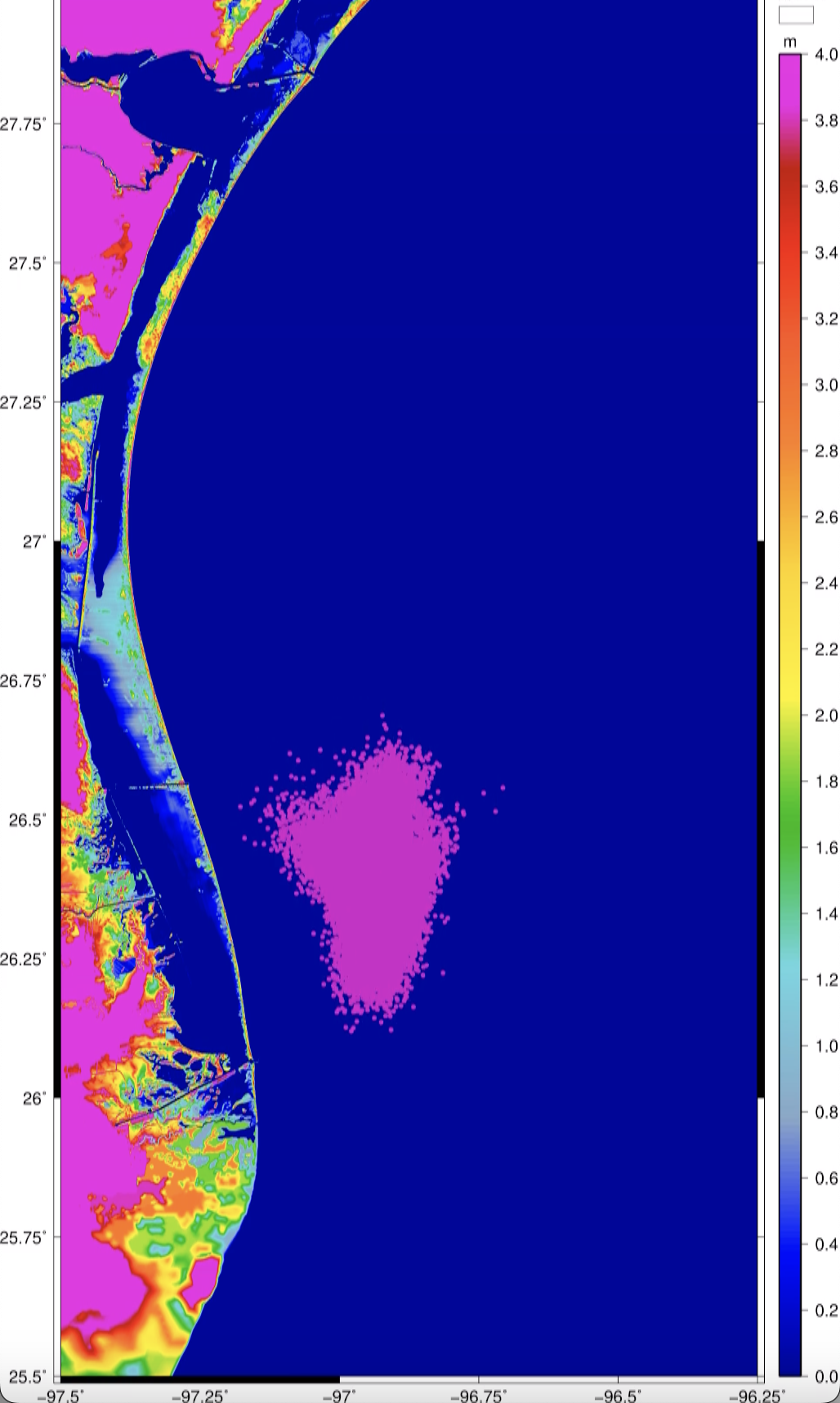}}
\caption{\label{fig:hurricane_nicholas} Particle locations during the moment of maximum spread using the hydrodynamic conditions of Hurricane Nicholas. The color spectrum denotes bathymetry above NAVD88 in meter.}
\end{figure}
In addition, oil particles remain entirely offshore when released at the offshore site and remain almost entirely in the coastal regions behind the barrier island when released from the onshore site, as seen in Figure~\ref{fig:hurricane_nicholas}.
%
% - harvey: actually a huge difference between existing and future for onshore
% - existing harvey onshore vs offshore: at the end of the 10 days, it looks like a similar-ish number of particles in the bays, but totally different trajectories throughout the 10 days
%
In contrast, the flow characteristics during the Category 4 Hurricane Harvey lead to vastly different particle trajectories between the current and proposed bathymetries. When particles are released from the onshore loading site, particles remain in the nearby bays for both mesh cases. However, using the mesh for the proposed bathymetry results in far more particles in the Corpus Christi Bay after ten days compared to the current bathymetry case (see Figure~\ref{fig:existing_harvey_onshore_10days} and Figure~\ref{fig:future_harvey_onshore_10days}). 
During these extreme flow conditions, a change in bathymetry leads to a more noticeable change in water velocities, which determine the particle trajectories. Figure~\ref{fig:maxvel} shows the difference in maximum velocity data from the two models using the two different meshes.
In the "normal" flow scenarios, particles released from the offshore site tend to remain offshore, whereas in the extreme hydrodynamic conditions seen during Hurricane Harvey, some particles released from the offshore site are transported to the coastline and into the nearby bays. It is important to note that the total number of particles that reach these bays is less than those that are released from the onshore location, as shown in Figure~\ref{fig:hurricane_harvey_bathymetry}.

\section{Concluding Remarks}
\label{sec:conclusions}

In this work, we have developed mathematical models governing the transport of oil particles from a proposed onshore loading terminal at Harbor Island, as well as at a proposed offshore loading site. The ADCIRC model is used establish flow velocity throughout the domain during various flow conditions, including the current and proposed channel depth, meteorological and tidal forcing during several different time periods, and two extreme weather events. To assess the extent of synthetic oil spills, oil particles are released in both onshore and offshore sites and their trajectories are tracked for one month in the normal flow cases and ten days during extreme weather events. The models consider the current  bathymetry of the Corpus Christi Ship Channel and the proposed bathymetry of a deepened ship channel.

Results from our numerical simulations suggest that bathymetry changes do not have a significant impact on the trajectories of oil particles during times of normal flow, but do affect particle trajectories during extreme flow conditions, such as those seen in Hurricane Harvey. While the spread of oil particles changes slightly in different seasons, the location of particle release (onshore or offshore loading sites) has the most significant effect on the particles being transported to the coastal ecosystems along the National Seashore and in the nearby bays. The extent of the spread is dependent on the time of year and presence of an extreme weather event.

Here, we have taken a commonly used approach to track the trajectory of oil during spills, i.e., modeling oil as distinct particles that are transported throughout a domain due to the flow of seawater. We have not considered chemical reactions and the potential degradation of oil as time progresses. It is also important to note that we do not take into account mitigation strategies to an oil spill, e.g., booms and the use of chemicals that disperses oil. We have only sought to investigate the potential transport of oil particles, should such spill be occur without human intervention. 

\subsubsection*{Acknowledgments}
\noindent The present study was conceived of and initiated by scientists at the University of Texas Marine Science Institute and carried out by students, faculty and staff from the Oden Institute for Computational Engineering and Sciences of the University of Texas at Austin. Funding provided by the Marine Science Institute included generous contributions made by members of the University of Texas Marine Science Advisory Council, in response to a special request for funds.
\\ The authors also would like to gratefully acknowledge the use of the ``ADCIRC", ``DMS23001", and ``DMS21031" allocations on the Frontera supercomputer at the Texas Advanced Computing Center at the University of Texas at Austin. 

\clearpage
\bibliography{sn-bibliography.bib}
%\printbibliography
\end{document}